\documentclass[a4paper,11pt]{article}
\pdfoutput=1 % if your are submitting a pdflatex (i.e. if you have
\usepackage{jheppub} % for details on the use of the package, please
\usepackage[T1]{fontenc} % if needed
\usepackage{cancel}
\usepackage{float}
\usepackage{graphicx}
\usepackage{subfig}
\usepackage{wasysym}
\usepackage[compat=1.1.0]{tikz-feynman}

\title{\boldmath Fermiophobic $Z'$ model for simultaneously explaining the muon anomalies $R_{K^{(*)}}$ and $(g-2)_{\mu}$}

\author[1]{Mario Fern\'andez Navarro}
\author[1]{and Stephen F. King}
\affiliation[1]{School of Physics \& Astronomy, University of Southampton, Southampton SO17 1BJ, UK}

\emailAdd{M.F.Navarro@soton.ac.uk}
\emailAdd{S.F.King@soton.ac.uk}

\abstract{We discuss a simple renormalisable, gauge invariant model with a fermiophobic $Z'$ boson: it has no couplings to the three Standard Model (SM)
chiral families, but does couple to a fourth vector-like (VL) family. The SM Higgs couples to the fourth VL lepton, leading
to an enhanced contribution to the muon anomalous magnetic moment
$\left(g-2\right)_{\mu}$. The latter contribution requires a non-vanishing
coupling of $Z'$ to right-handed muons, which arises within this model
due to mixing effects between the SM and VL fermions, along with $Z'$
couplings to the second generation SM lepton doublet and third
generation SM quark doublet. This model can simultaneously account
for the measured $B$-decay ratios $R_{K^{(*)}}$ and $\left(g-2\right)_{\mu}$. We identify the parameter
space where this explanation is consistent with existing experimental
constraints coming from $B_{s}-\bar{B}_{s}$ mixing, neutrino trident
production and collider searches. We also check that the SM Higgs
coupling to the fourth VL lepton does not produce a dangerous contribution
to the Higgs diphoton decay.}

\makeatletter
\gdef\@fpheader{}
\makeatother

\begin{document} 
\maketitle
\flushbottom

\section{Introduction}

Although the vast majority of particle-physics data is consistent
with the predictions of the Standard Model (SM), in recent times a
conspicuous series of discrepancies in flavour observables has been
established. One example is the discrepancy in rare flavour-changing
processes mediated by quark-level $b(\bar{b})\rightarrow s(\bar{s})\ell\bar{\ell}$
transitions, explored in the past by BaBar~\cite{BaBar:2012mrf} and
Belle~\cite{BELLE:2019xld}, along with LHC \cite{LHCb:2014vgu,LHCb:2019hip}.
In particular, the ratio of $B$-mesons decaying to $K\ell^{+}\ell^{-}$,
which involves a $\bar{b}\rightarrow\bar{s}\ell\bar{\ell}$ transition,
has been recently measured by LHCb \cite{LHCb:2021trn} in the dilepton
mass-squared range $1.1<q^{2}<6\,\mathrm{GeV}^{2}$ for the final
states $\mu^{+}\mu^{-}$ over $e^{+}e^{-}$,

\begin{equation}
R_{K}^{[1.1,6]}=\frac{\mathrm{Br}\left(B\rightarrow K\mu^{+}\mu^{-}\right)}{\mathrm{Br}\left(B\rightarrow Ke^{+}e^{-}\right)}=0.846_{-0.041}^{+0.044}\,,\label{eq:RK}
\end{equation}
along with the ratio of $B$-mesons decaying to $K^{*}\ell^{+}\ell^{-}$,
measured in the past by LHCb~\cite{LHCb:2017avl},

\begin{equation}
R_{K^{*}}^{[1.1,6]}=\frac{\mathrm{Br}\left(B\rightarrow K^{*}\mu^{+}\mu^{-}\right)}{\mathrm{Br}\left(B\rightarrow K^{*}e^{+}e^{-}\right)}=0.69_{-0.12}^{+0.16}\,.
\end{equation}

Within the SM, lepton universality predicts $R_{K^{(*)}}=1$, up to
corrections of order 1\% \cite{Descotes-Genon:2015uva,Bobeth:2007dw,Bordone:2016gaq,Straub:2018kue,Isidori:2020acz}
due to the different mass of muons and electrons. Hence, the previous
observations of $R_{K^{(*)}}$ seem to indicate the breaking of SM
lepton universality, up to the $3.1\sigma$ \cite{LHCb:2021trn} of
the most updated measurement of $R_{K}$, while $R_{K^{*}}$ is compatible
with the SM expectations at $2.4-2.5\sigma$~\cite{LHCb:2017avl}.

The apparent discrepancy of $R_{K^{(*)}}$ with the SM may be a hint
of new physics. Following these recent measurements of LHCb, a number
of phenomenological analyses of this data, see e.g.~Refs.~\cite{Hiller:2017bzc,Ciuchini:2017mik,Geng:2017svp,Capdevila:2017bsm,Ghosh:2017ber,Bardhan:2017xcc,Glashow:2014iga,DAmico:2017mtc,Calibbi:2015kma,Aebischer:2019mlg,Carvunis:2021jga,Angelescu:2021lln,Geng:2021nhg},
favour new physics operators of the form $\bar{s}_{L}\gamma_{\mu}b_{L}\bar{\mu}_{L}\gamma^{\mu}\mu_{L}$
or $\bar{s}_{L}\gamma_{\mu}b_{L}\bar{\mu}_{R}\gamma^{\mu}\mu_{R}$.
In particular, $R_{K^{(*)}}$ can be explained by only the purely
left-handed (LH) operator with a coefficient $\Lambda^{-2}$ where
$\Lambda\sim40\,\mathrm{TeV}$, or also by a linear combination of
both. Promising candidates for the arise of such effective operators
are tree-level exchange of a hypothetical, electrically neutral and
massive $Z'$ boson (see e.g.~\cite{Crivellin:2015mga,Crivellin:2015lwa,Chiang:2017hlj,King:2017anf,King:2018fcg,Falkowski:2018dsl})
with non-universal couplings to SM fermions, or the contribution of
a hypothetical leptoquark ($LQ$) coupling with different strengths
to the different types of charged leptons (see e.g.~\cite{Becirevic:2017jtw,deMedeirosVarzielas:2018bcy,DeMedeirosVarzielas:2019nob,King:2021jeo}).\\

Independent of the $R_{K^{(*)}}$ anomaly, there also exists a discrepancy
with the SM predictions in the experimentally measured anomalous magnetic
moments $a=(g-2)/2$ of both the muon and the electron. The long-lasting
non-compliance of the muon $a_{\mu}$ with the SM was first observed
by the Brookhaven E821 experiment at BNL \cite{Muong-2:2006rrc}.
This discrepancy has been recently confirmed by the most recent measurement
of the Fermilab experiment \cite{Muong-2:2021ojo},

\begin{equation}
\Delta a_{\mu}=a_{\mu}-a_{\mu}^{\mathrm{SM}}=\left(2.51\pm0.59\right)\cdot10^{-9}\,,\label{eq:Fermilab a_mu}
\end{equation}
a result $4.2\sigma$ greater than the SM prediction \cite{Aoyama:2020ynm,Aoyama:2012wk,Aoyama:2019ryr,Czarnecki:2002nt,Gnendiger:2013pva,Davier:2017zfy,Keshavarzi:2018mgv,Colangelo:2018mtw,Hoferichter:2019mqg,Davier:2019can,Keshavarzi:2019abf,Kurz:2014wya,Melnikov:2003xd,Masjuan:2017tvw,Colangelo:2017fiz,Hoferichter:2018kwz,Gerardin:2019vio,Bijnens:2019ghy,Colangelo:2019uex,Blum:2019ugy,Colangelo:2014qya}
and in excellent agreement with the previous BNL E821 measurement.
Such a discrepancy can also be addressed by $Z'$ models \cite{Altmannshofer:2016brv,CarcamoHernandez:2019ydc,Belanger:2015nma,CarcamoHernandez:2019xkb,Allanach:2015gkd,Raby:2017igl,Kawamura:2019rth,Kawamura:2019hxp,Kawamura:2021ygg}
or leptoquarks \cite{Cheung:2001ip,ColuccioLeskow:2016dox,Crivellin:2020tsz},
along with models involving extended scalar content and/or vector-like
(VL) fermions \cite{Arnan:2019uhr,Crivellin:2018qmi,Crivellin:2021rbq,Hernandez:2021tii}.
In particular, the minimal $Z'$ explanations \cite{Altmannshofer:2016brv} require
to introduce $\tau-\mu$ couplings in order to obtain an enhanced
contribution proportional to $m_{\tau}$. In such models, dangerous contributions to the flavour-violating processes $\tau\rightarrow3\mu$ or $\tau\rightarrow\mu\gamma$ may arise, along with possible breaking of lepton universality in leptonic tau decays, which is currently unobserved. Instead, Refs.~\cite{Belanger:2015nma,CarcamoHernandez:2019ydc,CarcamoHernandez:2019xkb,Allanach:2015gkd,Raby:2017igl,Kawamura:2019rth,Kawamura:2019hxp,Kawamura:2021ygg} consider a fermiophobic $Z'$ model where the $Z'$ couplings
with SM fermions are obtained through mixing with a fourth VL family.
An enhanced contribution to $\Delta a_{\mu}$ is obtained through
a coupling between the SM Higgs and a fourth VL lepton, although it has
to be checked that such a coupling would not spoil the existing Higgs
diphoton decay data. Moreover, this contribution requires a non-vanishing
coupling of $Z'$ to right-handed (RH) muons, in such a way that a purely
left-handed explanation of $R_{K^{(*)}}$, as in previous studies
\cite{King:2018fcg,Falkowski:2018dsl}, cannot be performed in this
case. Since the latest phenomenological analyses \cite{Angelescu:2021lln,Geng:2021nhg}
allow the possibility to include an effective operator $\bar{s}_{L}\gamma_{\mu}b_{L}\bar{\mu}_{R}\gamma^{\mu}\mu_{R}$
in the explanation of $R_{K^{(*)}}$, it could be possible to
simultaneously address $R_{K^{(*)}}$ and $\left(g-2\right)_{\mu}$ within
this fermiophobic $Z'$ framework.

However, it has to be checked whether such simultaneous explanation
of both anomalies can also preserve all currently released high energy
experimental data, such as the measurement of the mass difference
$\Delta M_{s}$ of neutral $B_{s}$ mesons, the observations of neutrino
trident production and the most recent collider signatures. Ideally,
such a model should be imminently testable with well designed future
searches. Moreover, $U(1)'$ extensions of the SM can be affected by Landau poles well below the Planck scale, and in some cases only a few orders of magnitude above the TeV scale \cite{Bause:2021prv}. However, we consider here a bottom-up approach, where the $U(1)'$ extension acts as an effective low energy theory, which would be embedded into a larger symmetry group below the energy scale of the Landau pole.

There are other $Z'$ models in the literature which address both
anomalies by considering a fourth VL family. In \cite{Belanger:2015nma}
the couplings to muons are loop-induced, while the model in \cite{CarcamoHernandez:2019xkb}
contains an extra $Z_{1}^{(1)}\times Z_{2}^{(2)}$ discrete symmetry
and the $Z'$ in \cite{Allanach:2015gkd,Raby:2017igl} is not fermiophobic. The models in~\cite{Kawamura:2019hxp,Kawamura:2019rth}
are similar model to that considered here but with general mixing between VL and
SM fermions, which leads to a large number of parameters, including all possible $Z'$ couplings to SM fermion, along with dangerous FCNCs and $Z-Z'$ kinetic mixing. Such a framework makes it difficult to systematically explore the parameter space, and instead a search of best fit points is performed. Moreover, such analyses reveal that the relevant parameters to simultaneously address $R_{K^{(*)}}$ and $\left(g-2\right)_{\mu}$ are only $Z'$ couplings to $bs$ quarks and muons. Hence, in contrast to the analyses in \cite{Kawamura:2019hxp,Kawamura:2019rth}, in the present paper we consider a 
simplified $Z'$ framework involving the fewest number of parameters in which the explanation of both anomalies can be simultaneously realised, allowing a systematic exploration of the parameter space.

The remainder of this article is organised as follows: in Section
\ref{sec:The-model} we outline the renormalisable and gauge invariant
fermiophobic model in which the $Z'$ only couples to a vector-like
fourth family. In Section \ref{sec:Mixing}, we show how it is possible
to switch on the couplings of the $Z'$ to the muon and $bs$-quarks
through mixing with the VL fermions, thereby eliminating all unnecessary
couplings and allowing us to focus on the connection between the $R_{K^{(*)}}$
and $\left(g-2\right)_{\mu}$ anomalies. The phenomenology and the
constraints that affect this model are presented in Section \ref{sec:Phenomenology-and-flavour}.
In Section \ref{sec:Results-and-discussion} we systematically explore
the parameter space of the model, and we also display and discuss
the results from our analysis. Finally, Section \ref{sec:Conclusions}
concludes the article.

\section{The model\label{sec:The-model}}
\begin{table}[!t]
\begin{centering}
\begin{tabular}{|ccccc|}
\hline 
\multicolumn{1}{|c|}{Field} & \multicolumn{1}{c|}{$SU(3)_{c}$} & \multicolumn{1}{c|}{$SU(2)_{L}$} & \multicolumn{1}{c|}{$U(1)_{Y}$} & $U(1)^{'}$\tabularnewline
\hline 
\hline 
$Q_{Li}=\left(\begin{array}{c}
u_{Li},\,\\
d_{Li}
\end{array}\right)$ & $\mathbf{3}$ & $\mathbf{2}$ & 1/6 & 0\tabularnewline
$u_{Ri}$ & $\mathbf{3}$ & $\mathbf{1}$ & 2/3 & 0\tabularnewline
$d_{Ri}$ & $\mathbf{3}$ & $\mathbf{1}$ & -1/3 & 0\tabularnewline
$L_{Li}=\left(\begin{array}{c}
\nu_{Li}\\
e_{Li}
\end{array}\right)$ & $\mathbf{1}$ & $\mathbf{2}$ & -1/2 & 0\tabularnewline
$e_{Ri}$ & $\mathbf{1}$ & $\mathbf{1}$ & -1 & 0\tabularnewline
\hline 
\noalign{\vskip\doublerulesep}
$Q_{L4},\widetilde{Q}_{R4}$ & $\mathbf{3}$ & $\mathbf{2}$ & 1/6 & $q_{Q_{4}}$\tabularnewline
$u_{R4},\widetilde{u}_{L4}$ & $\mathbf{3}$ & $\mathbf{1}$ & 2/3 & $q_{u_{4}}$\tabularnewline
$d_{R4},\widetilde{d}_{L4}$ & $\mathbf{3}$ & $\mathbf{1}$ & -1/3 & $q_{d_{4}}$\tabularnewline
$L_{L4},\widetilde{L}_{R4}$ & $\mathbf{1}$ & $\mathbf{2}$ & -1/2 & $q_{L_{4}}$\tabularnewline
$e_{R4},\widetilde{e}_{L4}$ & $\mathbf{1}$ & $\mathbf{1}$ & -1 & $q_{e_{4}}$\tabularnewline
$\nu_{R4},\widetilde{\nu}_{L4}$ & $\mathbf{1}$ & $\mathbf{1}$ & 0 & $q_{\nu_{4}}$\tabularnewline
\hline 
\hline 
$\phi_{f}$ & $\mathbf{1}$ & $\mathbf{1}$ & 0 & $-q_{f_{4}}$\tabularnewline
\hline 
\hline 
$H=\left(\begin{array}{c}
h^{+}\\
\left(v+h^{0}\right)/\sqrt{2}
\end{array}\right)$ & $\mathbf{1}$ & $\mathbf{2}$ & 1/2 & 0\tabularnewline
\hline 
\end{tabular}
\par\end{centering}
\caption{Particle assigments under $SU(3)_{c}\times SU(2)_{L}\times U(1)_{Y}\times U(1)'$
gauge symmetry, $i=1,2,3$. The singlet scalars $\phi_{f}$ ($f=Q,u,d,L,e,\nu$)
have $U(1)'$ charges $-q_{f_{4}}=-q_{Q_{4},u_{4},d_{4},L_{4},e_{4},\nu_{4}}$ \cite{King:2017anf}.
\label{tab:The-field-content}}
\end{table}
The model \cite{King:2017anf} (Table~\ref{tab:The-field-content}) includes the three
chiral families of LH $SU(2)_{L}$ doublets $\left(Q_{Li},\,L_{Li}\right)$
and RH $SU(2)_{L}$ singlets $\left(u_{Ri},\,d_{Ri},\,e_{Ri}\right)$ of the SM,
$i=1,2,3$; along with one vector-like family of fermions
(formed by LH and RH $SU(2)_{L}$ doublets $Q_{L4},\,L_{L4}$, $\text{and }\tilde{Q}_{R4},\,\tilde{L}_{R4}$,
together with LH and RH $SU(2)_{L}$ singlets $u_{R4},\,d_{R4},\,e_{R4},\,\nu_{R4}\text{ and }\widetilde{u}_{L4},\,\widetilde{d}_{L4},$
$\widetilde{e}_{L4},\,\widetilde{\nu}_{L4}$). The vector-like fermions
are charged under a gauge symmetry $U(1)'$, while the three chiral
families remain neutral under this symmetry, which is the reason behind
the model being called fermiophobic. The scalar sector is augmented
by gauge singlet fields $\phi_{f}$ with non-trivial charge assignments
$-q_{f_{4}}$ under the new symmetry, which are responsible for spontaneously
breaking $U(1)'$ developing vacuum expectation values
(VEVs) $\left\langle \phi_{f}\right\rangle $. The $Z'$ boson generated
after the symmetry breaking has a mass at the same scale $\left\langle \phi_{f}\right\rangle $.
\newpage
The full renormalisable Lagrangian is

\begin{align}
\begin{aligned}\mathcal{L}^{\mathrm{ren}} & =y_{ij}^{u}\overline{Q}_{Li}\widetilde{H}u_{Rj}+y_{ij}^{d}\overline{Q}_{Li}Hd_{Rj}+y_{ij}^{e}\overline{L}_{Li}He_{Rj} & {}\\
{} & +y_{4}^{u}\overline{Q}_{L4}\widetilde{H}u_{R4}+y_{4}^{d}\overline{Q}_{L4}Hd_{R4}+y_{4}^{e}\overline{L}_{L4}He_{R4}+y_{4}^{\nu}\overline{L}_{L4}\widetilde{H}\nu_{R4} & {}\\
{} & +x_{i}^{Q}\phi_{Q}\overline{Q}_{Li}\widetilde{Q}_{R4}+x_{i}^{L}\phi_{L}\overline{L}_{Li}\widetilde{L}_{R4}+x_{i}^{u}\phi_{u}^{*}\overline{\widetilde{u}}_{L4}u_{Ri}+x_{i}^{d}\phi_{d}^{*}\overline{\widetilde{d}}_{L4}d_{Ri}+x_{i}^{e}\phi_{e}^{*}\overline{\widetilde{e}}_{L4}e_{Ri} & {}\\
{} & +M_{4}^{Q}\overline{Q}_{L4}\widetilde{Q}_{R4}+M_{4}^{L}\overline{L}_{L4}\widetilde{L}_{R4}+M_{4}^{u}\overline{\widetilde{u}}_{L4}u_{R4}+M_{4}^{d}\overline{\widetilde{d}}_{L4}d_{R4}+M_{4}^{e}\overline{\widetilde{e}}_{L4}e_{R4}+M_{4}^{\nu}\overline{\widetilde{\nu}}_{L4}\nu_{R4}+\mathrm{h.c.} & {}
\end{aligned}
\label{eq: full lagrangian}
\end{align}
where $\widetilde{H}=i\sigma_{2}H^{*}$, $i=1,2,3$. The requirement
of $U(1)'$ invariance of the Yukawa interactions involving the fourth
family yields the following constraints on the $U(1)'$ charges:

\begin{equation}
q_{Q_{4}}=q_{u_{4}}=q_{d_{4}}\,,\qquad q_{L_{4}}=q_{e_{4}}=q_{\nu_{4}}\,.
\end{equation}

It is clear from Eq.~(\ref{eq: full lagrangian}) that fields in
the 4th, vector-like family obtain masses from two sources. Firstly,
from Yukawa terms involving the SM Higgs field, such as $y_{4}^{e}\overline{L}_{L4}He_{R4}$,
which get promoted to chirality-flipping fourth family mass terms
$M_{4}^{C}$ once the SM Higgs acquires a VEV. Secondly, from vector-like
mass terms, like $M_{4}^{L}\overline{L}_{L4}\widetilde{L}_{R4}$.
For the purpose of clarity, we shall treat $M_{4}^{C}$ and $M_{4}^{L}$
as independent masses in the analysis of the physical quantities of
interest, rather than constructing the full fourth family mass matrix
and diagonalising it, since such quantities rely on a chirality flip
and are sensitive to $M_{4}^{C}$ rather than the vector-like masses
$M_{4}^{L}$. Spontaneous breaking of $U(1)'$ by the scalars $\phi_{f}$
spontaneously acquiring VEVs gives rise to a massive $Z'$ boson featuring
couplings with the vector-like fermion fields. In the interaction
basis such terms will be diagonal and of the following form:
{\small{}
\begin{equation}
\mathcal{L}_{Z'}^{\mathrm{gauge}}=g'Z'_{\mu}\left(\overline{Q}_{L}D_{Q}\gamma^{\mu}Q_{L}+\overline{u}_{R}D_{u}\gamma^{\mu}u_{R}+\overline{d}_{R}D_{d}\gamma^{\mu}d_{R}+\overline{L}_{L}D_{L}\gamma^{\mu}L_{L}+\overline{e}_{R}D_{e}\gamma^{\mu}e_{R}+\overline{\nu}_{R}D_{\nu}\gamma^{\mu}\nu_{R}\right)\,,
\end{equation}
}{\small\par}

\begin{equation}
\begin{array}{c}
D_{Q}=\mathrm{diag}\left(0,0,0,q_{Q_{4}}\right)\,,\quad D_{u}=\mathrm{diag}\left(0,0,0,q_{Q_{4}}\right)\,,\quad D_{d}=\mathrm{diag}\left(0,0,0,q_{Q_{4}}\right)\,,\\
\,\\
D_{L}=\mathrm{diag}\left(0,0,0,q_{L_{4}}\right)\,,\quad D_{e}=\mathrm{diag}\left(0,0,0,q_{L_{4}}\right)\,,\quad D_{\nu}=\mathrm{diag}\left(0,0,0,q_{L_{4}}\right)\,.
\end{array}
\end{equation}

At this stage, the SM quarks and leptons do not couple to the $Z'$.
However, the Yukawa couplings detailed in Eq.~(\ref{eq: full lagrangian})
have no requirement to be diagonal. Before we can determine the full
masses of the propagating vector-like states and SM fermions, we need
to transform the field content of the model such that the Yukawa couplings
become diagonal. Therefore, fermions in the mass basis (denoted by
primed fields) are related to particles in the interaction basis by
the following unitary transformations

\begin{equation}
\begin{array}{c}
Q'_{L}=V_{Q_{L}}Q_{L}\,,\quad u'_{R}=V_{u_{R}}u_{R}\,,\quad d'_{R}=V_{d_{R}}d_{R}\,,\\
\,\\
L'_{L}=V_{L_{L}}L_{L}\,,\quad e'_{R}=V_{e_{R}}e_{R}\,,\quad\nu'_{R}=V_{\nu_{R}}\nu_{R}\,.
\end{array}
\end{equation}
This mixing induces couplings of SM mass eigenstate fermions to the
massive $Z'$, which can be expressed as follows

\begin{equation}
\begin{array}{c}
D'_{Q}=V_{Q_{L}}D_{Q}V_{Q_{L}}^{\dagger}\,,\quad D'_{u}=V_{u_{R}}D_{u}V_{u_{R}}^{\dagger}\,,\quad D'_{d}=V_{d_{R}}D_{d}V_{d_{R}}^{\dagger}\,,\\
\,\\
D'_{L}=V_{L_{L}}D_{L}V_{L_{L}}^{\dagger}\,,\quad D'_{e}=V_{e_{R}}D_{e}V_{e_{R}}^{\dagger}\,,\quad D'_{\nu}=V_{\nu_{R}}D_{\nu}V_{\nu_{R}}^{\dagger}\,.
\end{array}\label{eq:6}
\end{equation}

\section{Mixing\label{sec:Mixing}}

In this article, we consider a minimal mixing framework\footnote{Such a simplified mixing framework could be enforced by introducing some family symmetry, however a discussion of this is beyond the scope of this article.} in which both anomalies $R_{K^{(*)}}$ and $\left(g-2\right)_{\mu}$ can be simultaneously addressed. This requires that the fourth VL fermion family mixes only with the third generation of the SM quark doublet and with the second generation of the SM lepton doublet and singlet,

\begin{equation}
V_{Q_{L}}=V_{34}^{Q_{L}},\qquad V_{L_{L}}=V_{24}^{L_{L}},\qquad V_{e_{R}}=V_{24}^{e_{R}}\,,
\end{equation}
where

\begin{equation}
V_{34}^{Q_{L}}=\left(\begin{array}{cccc}
1 & 0 & 0 & 0\\
0 & 1 & 0 & 0\\
0 & 0 & \cos\theta_{34}^{Q} & \sin\theta_{34}^{Q}\\
0 & 0 & -\sin\theta_{34}^{Q} & \cos\theta_{34}^{Q}
\end{array}\right)\,,
\end{equation}

\begin{equation}
V_{24}^{L_{L},e_{R}}=\left(\begin{array}{cccc}
1 & 0 & 0 & 0\\
0 & \cos\theta_{24}^{L_{L},e_{R}} & 0 & \sin\theta_{24}^{L_{L},e_{R}}\\
0 & 0 & 1 & 0\\
0 & -\sin\theta_{24}^{L_{L},e_{R}} & 0 & \cos\theta_{24}^{L_{L},e_{R}}
\end{array}\right)\,,
\end{equation}
so for the matrices in Eq.~(\ref{eq:6}) we obtain

\begin{equation}
D'_{Q}=V_{34}^{Q_{L}}D_{Q}\left(V_{34}^{Q_{L}}\right)^{\dagger}=q_{Q_{4}}\left(\begin{array}{cccc}
0 & 0 & 0 & 0\\
0 & 0 & 0 & 0\\
0 & 0 & \left(\sin\theta_{34}^{Q}\right)^{2} & \cos\theta_{34}^{Q}\sin\theta_{34}^{Q}\\
0 & 0 & \cos\theta_{34}^{Q}\sin\theta_{34}^{Q} & \left(\cos\theta_{34}^{Q}\right)^{2}
\end{array}\right)\,,
\end{equation}

\begin{equation}
D'_{L}=V_{24}^{L_{L}}D_{L}\left(V_{24}^{L_{L}}\right)^{\dagger}=q_{L_{4}}\left(\begin{array}{cccc}
0 & 0 & 0 & 0\\
0 & \left(\sin\theta_{24}^{L_{L}}\right)^{2} & 0 & \cos\theta_{24}^{L_{L}}\sin\theta_{24}^{L_{L}}\\
0 & 0 & 0 & 0\\
0 & \cos\theta_{24}^{L_{L}}\sin\theta_{24}^{L_{L}} & 0 & \left(\cos\theta_{24}^{L_{L}}\right)^{2}
\end{array}\right)\,,
\end{equation}

\begin{equation}
D'_{e}=V_{24}^{e_{R}}D_{e}\left(V_{24}^{e_{R}}\right)^{\dagger}=q_{L_{4}}\left(\begin{array}{cccc}
0 & 0 & 0 & 0\\
0 & \left(\sin\theta_{24}^{e_{R}}\right)^{2} & 0 & \cos\theta_{24}^{e_{R}}\sin\theta_{24}^{e_{R}}\\
0 & 0 & 0 & 0\\
0 & \cos\theta_{24}^{e_{R}}\sin\theta_{24}^{e_{R}} & 0 & \left(\cos\theta_{24}^{e_{R}}\right)^{2}
\end{array}\right)\,,
\end{equation}
hence in this basis the relevant $Z'$ couplings read

\begin{equation}
\mathcal{L}_{Z'}\supset Z'_{\mu}\left(g_{bb}\overline{b}_{L}\gamma^{\mu}b_{L}+g_{\mu\mu}^{L}\overline{\mu}_{L}\gamma^{\mu}\mu_{L}+g_{\mu\mu}^{R}\overline{\mu}_{R}\gamma^{\mu}\mu_{R}\right)\,,
\end{equation}
where

\begin{equation}
g_{bb}=g'q_{Q_{4}}\left(\sin\theta_{34}^{Q}\right)^{2}\,,
\end{equation}

\begin{equation}
g_{\mu\mu}^{L}=g'q_{L_{4}}\left(\sin\theta_{24}^{L_{L}}\right)^{2}\,,\label{eq:gLmumu}
\end{equation}

\begin{equation}
g_{\mu\mu}^{R}=g'q_{L_{4}}\left(\sin\theta_{24}^{e_{R}}\right)^{2}\,.\label{eq:gRmumu}
\end{equation}
We also obtain a CKM suppressed $bs$ coupling in the basis in which the up-quark mass matrix is diagonal. 
In this basis, $V_{\mathrm{CKM}}=~V^{\dagger}_{d_{L}}$, and we find the couplings

\begin{equation}
Z'g_{bs}\bar{s}_{L}\gamma^{\mu}b_{L}\,,\label{eq:bs coupling}
\end{equation}

\begin{equation}
g_{bs}=g_{bb}V_{ts}=g'q_{Q_{4}}\left(\sin\theta_{34}^{Q}\right)^{2}V_{ts}\,,
\end{equation}
where $V_{ts}\approx-0.04$. Usually $R_{K^{(*)}}$ can be addressed with just $g_{bs}$ and $g_{\mu\mu}^{L}$ couplings (see e.g.~\cite{King:2017anf,King:2018fcg,Falkowski:2018dsl}), but we also need $g_{\mu\mu}^{R}$ in order to simultanously explain $\left(g-2\right)_{\mu}$ in this model, as we shall see in the next section.

\section{Phenomenology and constraints\label{sec:Phenomenology-and-flavour}}

\subsection{$(g-2)_{\mu}$}

\begin{figure}[H]
$\qquad$\subfloat[]{\begin{centering}
\begin{tikzpicture}
	\begin{feynman}
		\vertex (a) {\(\mu_{R,L}\)};
		\vertex [right=18mm of a] (b);
		\vertex [right=of b] (c);
		\vertex [right=of c] (d){\(\mu_{L,R}\)};
		\diagram* {
			(a) -- [edge label'=\(\mu_{L,R}\), near end, inner sep=6pt, insertion=0.5] (b) -- [boson, half left, edge label=$Z'$] (c),
			(b) --  [fermion, edge label'=\(\mu_{L,R}\), inner sep=6pt] (c),
			(c) -- [fermion] (d),
	};
	\end{feynman}
\end{tikzpicture}
\par\end{centering}
}$\qquad\quad\qquad$\subfloat[]{\begin{centering}
\begin{tikzpicture}
	\begin{feynman}
		\vertex (a) {\(\mu_{R,L}\)};
		\vertex [right=18mm of a] (b);
		\vertex [right=of b] (c);
		\vertex [right=of c] (d){\(\mu_{L,R}\)};
		\diagram* {
			(a) -- [edge label'=\(\mu_{L,R}\), near end, inner sep=6pt, insertion=0.5] (b) -- [boson, half left, edge label=$Z'$] (c),
			(b) --  [fermion, edge label'=\(E_{4L,R}\), inner sep=6pt] (c),
			(c) -- [fermion] (d),
	};
	\end{feynman}
\end{tikzpicture}
\par\end{centering}
}

\subfloat[]{\begin{centering}
\begin{tikzpicture}
	\begin{feynman}
		\vertex (a) {\(\mu_{R,L}\)};
		\vertex [right=18mm of a] (b);
		\vertex [right=of b] (c) [label={ [xshift=0.1cm, yshift=0.1cm] \small $m_{\mu}$}];
		\vertex [right=of c] (d);
		\vertex [right=of d] (e) {\(\mu_{L,R}\)};
		\diagram* {
			(a) -- [fermion] (b) -- [boson, half left, edge label=$Z'$] (d),
			(b) -- [edge label'=\(\mu_{R,L}\)] (c),
			(c) -- [edge label'=\(\mu_{L,R}\), inner sep=4pt, insertion=0] (d),
			(d) -- [fermion] (e),
	};
	\end{feynman}
\end{tikzpicture}
\par\end{centering}
}$\quad$\subfloat[]{\begin{centering}
\begin{tikzpicture}
	\begin{feynman}
		\vertex (a) {\(\mu_{R,L}\)};
		\vertex [right=18mm of a] (b);
		\vertex [right=of b] (c) [label={ [xshift=0.1cm, yshift=0.1cm] \small $M^{C}_{4}$}];
		\vertex [right=of c] (d);
		\vertex [right=of d] (e) {\(\mu_{L,R}\)};
		\diagram* {
			(a) -- [fermion] (b) -- [boson, half left, edge label=$Z'$] (d),
			(b) -- [edge label'=\(E_{4R,L}\)] (c),
			(c) -- [edge label'=\(E_{4L,R}\), inner sep=4pt, insertion=0] (d),
			(d) -- [fermion] (e),
	};
	\end{feynman}
\end{tikzpicture}
\par\end{centering}
}\caption{Feynman diagrams in the model contributing to $\left(g-2\right)_{\mu}$, photon lines are implicit.
\label{fig: g-2 diagrams}}
\end{figure}
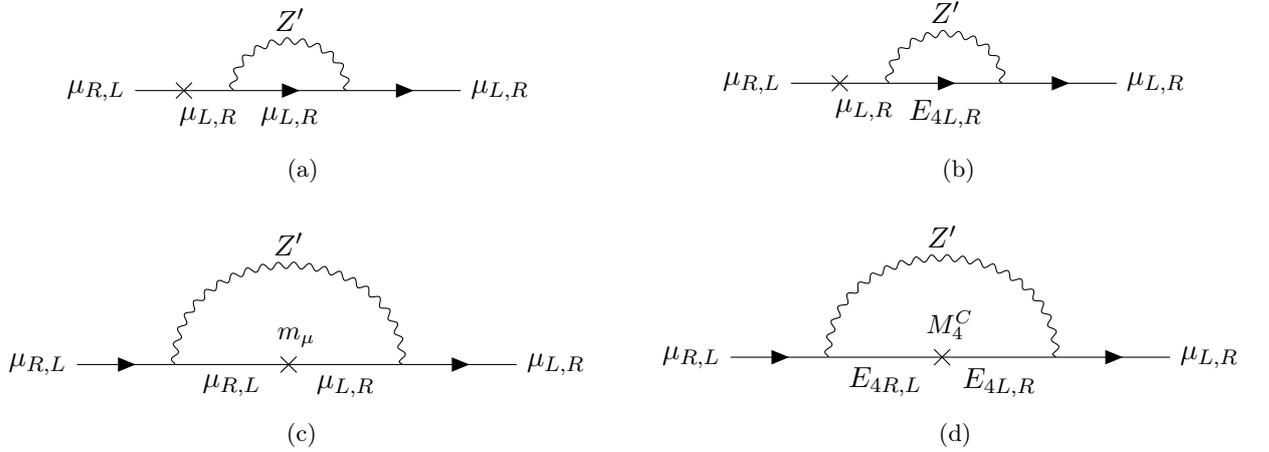

The diagrams displayed in Fig.~\ref{fig: g-2 diagrams} lead to $Z'$-mediated
contributions to the muon anomalous magnetic moment, namely \cite{CarcamoHernandez:2019ydc}

\begin{equation}
\begin{aligned}\Delta a_{\mu}=-\frac{m_{\mu}^{2}}{8\pi^{2}M_{Z'}^{2}}\left[\vphantom{\frac{A^{C}}{A_{\mu}}}\right. & \left(\left|g_{\mu\mu}^{L}\right|^{2}+\left|g_{\mu\mu}^{R}\right|^{2}\right)F(m_{\mu}^{2}/M_{Z'}^{2})+\left(\left|g_{\mu E}^{L}\right|^{2}+\left|g_{\mu E}^{R}\right|^{2}\right)F(m_{E}^{2}/M_{Z'}^{2}) & {}\\
{} & {\displaystyle \left.+\mathrm{Re}\left[g_{\mu\mu}^{L}\left(g_{\mu\mu}^{R}\right)^{*}\right]G(m_{\mu}^{2}/M_{Z'}^{2})+\mathrm{Re}\left[g_{\mu E}^{L}\left(g_{\mu E}^{R}\right)^{*}\right]\frac{M_{4}^{C}}{m_{\mu}}G(m_{E}^{2}/M_{Z'}^{2})\right]\,,} & {}
\end{aligned}
\label{eq:g-2}
\end{equation}
where $G(x)$ and $F(x)$ are $\mathcal{O}(1)$ loop functions, and
$m_{E}$ is the propagating mass of the 4th lepton. In our case, $m_{E}\simeq M_{4}^{L}$
since we consider that the dominant source of mass for the 4th lepton
is vector-like, i.e. $M_{4}^{L}\gg M_{4}^{C}$. For the upcoming sections
we shall fix $M_{4}^{L}=5\,\mathrm{TeV}$, in order to preserve $M_{4}^{L}\gg M_{4}^{C}$
for a chirality-flipping mass $M_{4}^{C}$ of order GeV. The couplings
between muons and VL leptons read

\begin{equation}
g_{\mu E}^{L}=g'q_{L4}\cos\theta_{24}^{L_{L}}\sin\theta_{24}^{L_{L}}=g'q_{L4}\sqrt{1-g_{\mu\mu}^{L}/(g'q_{L4})}\sqrt{g_{\mu\mu}^{L}/(g'q_{L4})}\,,
\end{equation}

\begin{equation}
g_{\mu E}^{R}=g'q_{L4}\cos\theta_{24}^{e_{R}}\sin\theta_{24}^{e_{R}}=g'q_{L4}\sqrt{1-g_{\mu\mu}^{R}/(g'q_{L4})}\sqrt{g_{\mu\mu}^{R}/(g'q_{L4})}\,,
\end{equation}
where from now on we will assume $g'q_{L4}=1$ for simplicity.\\

Since the loop functions satisfy $G(x)<0$ and $F(x)>0$, the contributions
proportional to $G(x)$ and $F(x)$ in Eq.~(\ref{eq:g-2}) interfere
negatively. However, for a chirality-flipping mass $M_{4}^{C}$ of
order $v/\sqrt{2}$ (where $v=246\,\mathrm{GeV}$ is the SM Higgs
VEV), the term proportional to $M_{4}^{C}$ in Eq.~(\ref{eq:g-2})
is dominant and positive due to $G(x)<0$, matching the required sign
to explain the experimental measurement of $\Delta a_{\mu}$ by Fermilab
\cite{Muong-2:2021ojo} (see Eq.~(\ref{eq:Fermilab a_mu})). Hence,
a non-vanishing coupling of $Z'$ to RH muons is crucial to explain
$(g-2)_{\mu}$ here: otherwise, if we assume $g_{\mu\mu}^{R}=0$,
then $g_{\mu E}^{R}$ vanishes and we lose the dominant contribution
proportional to $M_{4}^{C}$.

\subsection{$R_{K^{(*)}}$}

\begin{figure}[H]
\subfloat[$Z'$ exchange diagrams contributing to $R_{K^{(*)}}$. \label{fig: Z' exchange to R_K}]{\begin{tikzpicture}
	\begin{feynman}
		\vertex (a) {\(b_{L}\)};
		\vertex [right=18mm of a] (b);
		\vertex [below right=of b] (c);
		\vertex [above right=of b] (f1) {\(s_{L}\)};
		\vertex [above right=of c] (f2) {\(\mu_{L,R}\)};
		\vertex [below right=of c] (f3) {\(\bar{\mu}_{L,R}\)};
		\diagram* {
			(a) -- [fermion] (b) -- [fermion] (f1),
			(b) -- [boson, blue, edge label'=\(Z'\), inner sep=6pt] (c),
			(c) -- [fermion] (f2),
			(c) -- [anti fermion] (f3),
	};
	\end{feynman}
\end{tikzpicture}

}$\qquad\qquad\qquad\qquad$\subfloat[$Z'$ exchange diagrams contributing to neutrino trident production.
\label{fig: Z' exchange to trident}]{\begin{tikzpicture}
	\begin{feynman}
		\vertex (a) {\(\nu_{\mu L}\)};
		\vertex [right=18mm of a] (b);
		\vertex [below right=of b] (c);
		\vertex [above right=of b] (f1) {\(\nu_{\mu L}\)};
		\vertex [above right=of c] (f2) {\(\mu_{L,R}\)};
		\vertex [below right=of c] (f3) {\(\bar{\mu}_{L,R}\)};
		\diagram* {
			(a) -- [fermion] (b) -- [fermion] (f1),
			(b) -- [boson, blue, edge label'=\(Z'\), inner sep=6pt] (c),
			(c) -- [fermion] (f2),
			(c) -- [anti fermion] (f3),
	};
	\end{feynman}
\end{tikzpicture}

}

\subfloat[$Z'$ exchange diagrams contributing to $B_{s}\rightarrow\bar{\mu}\mu$.
\label{fig: Z' exchange to Bsmumu}]{\begin{tikzpicture}
	\begin{feynman}
		\vertex (a) {\(\bar{s}_{L}\)};
		\vertex [above right=21mm of a] (b);
		\vertex [right=of b] (c);
		\vertex [above left=of b] (f1) {\(b_{L}\)};
		\vertex [above right=of c] (f2) {\(\mu_{L,R}\)};
		\vertex [below right=of c] (f3) {\(\bar{\mu}_{L,R}\)};
		\diagram* {
			(a) -- [anti fermion] (b) -- [anti fermion] (f1),
			(b) -- [boson, blue, edge label'=\(Z'\), inner sep=6pt] (c),
			(c) -- [fermion] (f2),
			(c) -- [anti fermion] (f3),
	};
	\end{feynman}
\end{tikzpicture}

}$\qquad\qquad\qquad\qquad$\subfloat[$Z'$ exchange diagrams contributing to $B_{s}-\bar{B}_{s}$ mixing.
\label{fig: Z' exchange to Bs mixing}]{\begin{tikzpicture}
	\begin{feynman}
		\vertex (a) {\(\bar{s}_{L}\)};
		\vertex [above right=21mm of a] (b);
		\vertex [right=of b] (c);
		\vertex [above left=of b] (f1) {\(b_{L}\)};
		\vertex [above right=of c] (f2) {\(s_{L}\)};
		\vertex [below right=of c] (f3) {\(\bar{b}_{L}\)};
		\diagram* {
			(a) -- [anti fermion] (b) -- [anti fermion] (f1),
			(b) -- [boson, blue, edge label'=\(Z'\), inner sep=6pt] (c),
			(c) -- [fermion] (f2),
			(c) -- [anti fermion] (f3),
	};
	\end{feynman}
\end{tikzpicture}

}\caption{}
\end{figure}

One possible explanation of the $R_{K^{(*)}}$ measurements in LHCb
is that the low-energy Lagrangian below the EW scale contains additional
contributions to the effective 4-fermion operator with left/right-handed
muon, left-handed $b$-quark, and left-handed $s$-quark fields,

\begin{equation}
\Delta\mathcal{L}_{\mathbf{\mathrm{eff}}}\supset G_{bs\mu}^{L}\left(\bar{s}_{L}\gamma_{\mu}b_{L}\right)\left(\bar{\mu}_{L}\gamma^{\mu}\mu_{L}\right)+G_{bs\mu}^{R}\left(\bar{s}_{L}\gamma_{\mu}b_{L}\right)\left(\bar{\mu}_{R}\gamma^{\mu}\mu_{R}\right)+\mathrm{h.c.}\,,
\end{equation}
arising in our model from integrating out the $Z'$ boson at tree-level (Fig.~\ref{fig: Z' exchange to R_K}). The above operators contribute to the flavour changing transitions
$b_{L}\rightarrow s_{L}\bar{\mu}_{L}\mu_{L}$ and $b_{L}\rightarrow s_{L}\bar{\mu}_{R}\mu_{R}$,
respectively. A $Z'$-mediated contribution to $B_{s}\rightarrow\bar{\mu}\mu$
(Fig.~\ref{fig: Z' exchange to Bsmumu}) also arises.

We can express the coefficients $G_{bs\mu}^{L}$ and $G_{bs\mu}^{R}$
as a function of the couplings $g_{bb}$, $g_{\mu\mu}^{L}$ and $g_{\mu\mu}^{R}$,

\begin{equation}
G_{bs\mu}^{L}=-\frac{V_{ts}g_{bb}g_{\mu\mu}^{L}}{M_{Z'}^{2}}=\frac{-V_{ts}\left(g'\right)^{2}q_{Q_{4}}q_{L4}\left(\sin\theta_{34}^{Q}\right)^{2}\left(\sin\theta_{24}^{L_{L}}\right)^{2}}{M_{Z'}^{2}}\,,\label{eq:GLbsmu}
\end{equation}

\begin{equation}
G_{bs\mu}^{R}=-\frac{V_{ts}g_{bb}g_{\mu\mu}^{R}}{M_{Z'}^{2}}=\frac{-V_{ts}\left(g'\right)^{2}q_{Q_{4}}q_{L4}\left(\sin\theta_{34}^{Q}\right)^{2}\left(\sin\theta_{24}^{e_{R}}\right)^{2}}{M_{Z'}^{2}}\,,\label{eq:GRbsmu}
\end{equation}
where it can be seen that both $G_{bs\mu}^{L}$ and $G_{bs\mu}^{R}$
have the same sign in our model.

In Ref.~\cite{Geng:2021nhg}, the vector and axial effective operators

\begin{equation}
\mathcal{H}_{\mathrm{eff}}\supset\mathcal{N}\left[\delta C_{9}\left(\bar{s}_{L}\gamma_{\mu}b_{L}\right)\left(\bar{\mu}\gamma^{\mu}\mu\right)+\delta C_{10}\left(\bar{s}_{L}\gamma_{\mu}b_{L}\right)\left(\bar{\mu}\gamma^{\mu}\gamma_{5}\mu\right)\right]+\mathrm{h.c.}\,,
\end{equation}

\begin{equation}
\mathcal{N}=-\frac{4G_{\mathrm{F}}}{\sqrt{2}}V_{tb}V_{ts}^{*}\frac{e^{2}}{16\pi^{2}}\,,
\end{equation}
had been fitted to explain $R_{K^{(*)}}$ up to the $1\sigma$ level,
as shown in Tables~\ref{tab:Theoretically-clean-fit} and \ref{tab:Fit-including-angular}.
From the results for $\delta C_{9}$ and $\delta C_{10}$ we have
computed the numerical values of $G_{bs\mu}^{L}$ and $G_{bs\mu}^{R}$
that fit $R_{K^{(*)}}$ up to the 1$\sigma$ level,

\begin{equation}
\delta C_{9}=-\frac{G_{bs\mu}^{L}+G_{bs\mu}^{R}}{2\mathcal{N}}\Rightarrow\,G_{bs\mu}^{L}=\mathcal{N}\left(\delta C_{10}-\delta C_{9}\right)\,,
\end{equation}

\begin{equation}
\delta C_{10}=\frac{G_{bs\mu}^{L}-G_{bs\mu}^{R}}{2\mathcal{N}}\Rightarrow\,G_{bs\mu}^{R}=-\mathcal{N}\left(\delta C_{9}+\delta C_{10}\right)\,.
\end{equation}

The results displayed in Table~\ref{tab:Theoretically-clean-fit}
consider the so-called ``\textbf{theoretically clean} \textbf{fit}''
which, as explained in Ref.~\cite{Geng:2021nhg}, displays the values
of $G_{bs\mu}^{L}$ and $G_{bs\mu}^{R}$ that simultaneously fit $R_{K^{(*)}}$
and the $B_{s}\rightarrow\bar{\mu}\mu$ data. This fit is denoted as theoretically
clean since all the observables included are free from theoretical
uncertainties. On the other hand, the \textbf{global fit} in Table~\ref{tab:Fit-including-angular}
also includes the fit of angular observables in $B\rightarrow K^{*}\bar{\mu}\mu$
data reported by LHCb, ATLAS and CMS, which are afflicted by larger
theoretical uncertainties than the ratios of lepton  universality violation
and the $B_{s}\rightarrow\bar{\mu}\mu$ data \cite{Geng:2021nhg}. 
\begin{center}
\begin{table}[H]
\caption{Fit of $R_{K^{(*)}}$ and the $B_{s}\rightarrow\bar{\mu}\mu$ data\textbf{
(Theoretically Clean Fit) }\cite{Geng:2021nhg} \label{tab:Theoretically-clean-fit}}

\centering{}{\small{}}%
\begin{tabular}{|c|c|c|}
\hline 
 & {\small{}Best fit } & {\small{}$1\sigma$ range}\tabularnewline
\hline 
\hline 
{\small{}$\left(\delta C_{9},\delta C_{10}\right)$ } & {\small{}$\left(-0.11,0.59\right)$ } & {\small{}$\begin{array}{c}
\delta C_{9}\mathcal{2}\left[-0.41,\,0.17\right],\end{array}\delta C_{10}\mathcal{2}\left[0.38,\,0.81\right]$ }\tabularnewline
\hline 
{\small{}$\left(G_{bs\mu}^{L}/\mathcal{N},G_{bs\mu}^{R}/\mathcal{N}\right)$} & {\small{}$\left(0.7,-0.48\right)$ } & {\small{}$\begin{array}{c}
G_{bs\mu}^{L}/\mathcal{N}\mathcal{2}\left[0.64,\,0.79\right],G_{bs\mu}^{R}/\mathcal{N}\mathcal{2}\left[-0.98,\,0.03\right]\end{array}$}\tabularnewline
\hline 
{\small{}$\left(G_{bs\mu}^{L},G_{bs\mu}^{R}\right)$ } & {\small{}${\displaystyle \left(\frac{1}{\left(42.5\,\mathrm{TeV}\right)^{2}},\,-\frac{1}{\left(51.3\,\mathrm{TeV}\right)^{2}}\right)}$ } & {\small{}$\begin{array}{c}
{\displaystyle \,}\\
{\displaystyle G_{bs\mu}^{L}\mathcal{2}\left[\frac{1}{\left(44.44\,\mathrm{TeV}\right)^{2}},\,\frac{1}{\left(40\,\mathrm{TeV}\right)^{2}}\right],}\\
\,\\
{\displaystyle G_{bs\mu}^{R}\mathcal{2}\left[-\frac{1}{\left(35.9\,\mathrm{TeV}\right)^{2}},\,\frac{1}{\left(205\,\mathrm{TeV}\right)^{2}}\right]}\\
\,
\end{array}$}\tabularnewline
\hline 
\end{tabular}{\small\par}
\end{table}
\par\end{center}

\begin{center}
\begin{table}[H]
\caption{Fit of $R_{K^{(*)}}$, $B_{s}\rightarrow\bar{\mu}\mu$ data and angular
observables of $B\rightarrow K^{*}\bar{\mu}\mu$ data (\textbf{Global
Fit}) \cite{Geng:2021nhg}\label{tab:Fit-including-angular}}

\centering{}{\small{}}%
\begin{tabular}{|c|c|c|}
\hline 
 & {\small{}Best fit } & {\small{}$1\sigma$ range}\tabularnewline
\hline 
\hline 
{\small{}$\left(\delta C_{9},\delta C_{10}\right)$ } & {\small{}$\left(-0.56,0.30\right)$ } & {\small{}$\begin{array}{c}
\delta C_{9}\mathcal{2}\left[-0.79,\,-0.31\right],\delta C_{10}\mathcal{2}\left[0.15,\,0.49\right]\end{array}$}\tabularnewline
\hline 
{\small{}$\left(G_{bs\mu}^{L}/\mathcal{N},G_{bs\mu}^{R}/\mathcal{N}\right)$} & {\small{}$\left(0.86,0.26\right)$ } & {\small{}$\begin{array}{c}
G_{bs\mu}^{L}/\mathcal{N}\mathcal{2}\left[0.8,\,0.94\right],G_{bs\mu}^{R}/\mathcal{N}\mathcal{2}\left[-0.18,\,0.64\right]\end{array}$ }\tabularnewline
\hline 
{\small{}$\left(G_{bs\mu}^{L},G_{bs\mu}^{R}\right)$ } & {\small{}${\displaystyle \left(\frac{1}{\left(38.34\,\mathrm{TeV}\right)^{2}},\,\frac{1}{\left(69.73\,\mathrm{TeV}\right)^{2}}\right)}$ } & {\small{}$\begin{array}{c}
{\displaystyle \,}\\
{\displaystyle G_{bs\mu}^{L}\mathcal{2}\left[\frac{1}{\left(39.75\,\mathrm{TeV}\right)^{2}},\,\frac{1}{\left(36.67\,\mathrm{TeV}\right)^{2}}\right],}\\
\,\\
{\displaystyle G_{bs\mu}^{R}\mathcal{2}\left[-\frac{1}{\left(83.8\,\mathrm{TeV}\right)^{2}},\,\frac{1}{\left(44.44\,\mathrm{TeV}\right)^{2}}\right]}\\
\,
\end{array}$}\tabularnewline
\hline 
\end{tabular}{\small\par}
\end{table}
\par\end{center}

On one hand, $G_{bs\mu}^{L}$ shows similar best fit values of order $(40$ $\mathrm{TeV)^{-2}}$ in both
fits, although the $1\sigma$ region is slightly tighter in the global fit
(Table~\ref{tab:Fit-including-angular}) than in the theoretically
clean fit (Table~\ref{tab:Theoretically-clean-fit}). On the other hand,
$G_{bs\mu}^{R}$ shows the largest differences between both fits.
For the theoretically clean fit, $G_{bs\mu}^{R}<0$ is favoured, although
$G_{bs\mu}^{R}>0$ is still allowed. For the global fit, the situation
is the opposite: $G_{bs\mu}^{R}>0$ is favoured, although $G_{bs\mu}^{R}<0$
is also allowed. As a consequence, in both fits $G_{bs\mu}^{R}$ is
compatible with zero and hence $R_{K^{(*)}}$ can also be explained
with only the purely left-handed operator $\bar{s}_{L}\gamma_{\mu}b_{L}\bar{\mu}_{L}\gamma^{\mu}\mu_{L}$,
as in previous $Z'$ models \cite{Falkowski:2018dsl,King:2018fcg,King:2017anf}.
However, we have shown that we need a non-vanishing coupling of right-handed muons
to $Z'$ in order to explain $\left(g-2\right)_{\mu}$, hence within
this model we have a non-zero right-handed contribution to $R_{K^{(*)}}$. Therefore,
we need to be aware of keeping such contribution, i.e. $G_{bs\mu}^{R}$,
within the $1\sigma$ region of the considered fit. \\

Moreover, the best fit value of $G_{bs\mu}^{R}$ is negative within
the theoretically clean fit, but positive within the global fit. This
indicates that the extra angular observables of ${B\rightarrow K^{*}\bar{\mu}\mu}$
data are relevant and drastically change the picture for explaining
$R_{K^{(*)}}$ with effective operators $\bar{s}_{L}\gamma_{\mu}b_{L}\bar{\mu}_{R}\gamma^{\mu}\mu_{R}$.
However, the fact that these angular observables are affected by important
theoretical uncertainties lead to some tension in the community about
whether angular observables of ${B\rightarrow K^{*}\bar{\mu}\mu}$ data should
be considered or not in the global fits. Because of this, during the remainder of this
work we will consider both fits for computing our results. On the
other hand, in our model $G_{bs\mu}^{L}$ and $G_{bs\mu}^{R}$ must
have the same relative sign. Therefore, we shall keep the product
$q_{Q_{4}}q_{L4}$ positive and then fit $G_{bs\mu}^{R}$ in the positive
region allowed within the $1\sigma$. We shall study whether this
can be challenging in the theoretically clean fit, where the positive
region allowed by the $1\sigma$ range of $G_{bs\mu}^{R}$ is tiny.
In other words, the requirement of keeping $G_{bs\mu}^{R}$ within
the $1\sigma$ range of the theoretically clean fit constitutes an
extra effective constraint over this model.

\subsection{$B_{s}-\bar{B}_{s}$ mixing}

The $Z'$ coupling to $bs$-quarks in Eq.~(\ref{eq:bs coupling}) leads
to an additional tree-level contribution (Fig.~\ref{fig: Z' exchange to Bs mixing})
to $B_{s}-\bar{B}_{s}$ mixing,

\begin{equation}
\Delta\mathcal{L}_{\mathbf{\mathrm{eff}}}\supset-\frac{G_{bs}}{2}\left(\bar{s}_{L}\gamma^{\mu}b_{L}\right)^{2}+\mathrm{h.c.}
\end{equation}
where

\begin{equation}
G_{bs}=\frac{g_{bs}^{2}}{M_{Z'}^{2}}=\frac{g_{bb}^{2}V_{ts}^{2}}{M_{Z'}^{2}}\,.
\end{equation}

Such a new contribution is constrained by the results of the mass
difference $\Delta M_{s}$ of neutral $B_{s}$ mesons. The theoretical
determination of the mass difference is limited by our understanding
of non-perturbative matrix elements of dimension six operators, which
can be computed with lattice simulations or sum rules. Here we follow
the recent analysis of Ref.~\cite{DiLuzio:2019jyq}, which displays
two different results for $\Delta M_{s}$,

\begin{equation}
\Delta M_{s}^{\mathrm{FLAG'19}}=\left(1.13_{-0.07}^{+0.07}\right)\Delta M_{s}^{\mathrm{exp}}\,,
\end{equation}

\begin{equation}
\Delta M_{s}^{\mathrm{Average'19}}=\left(1.04_{-0.09}^{+0.04}\right)\Delta M_{s}^{\mathrm{exp}}\,.
\end{equation}
$\Delta M_{s}^{\mathrm{FLAG'19}}$ is obtained using lattice results,
and is about two standard deviations above the experimental numbers.
This result for the mass difference sets the strong bound

\begin{equation}
G_{bs}\lesssim\frac{1}{\left(330\,\mathrm{TeV}\right)^{2}}\,.
\end{equation}
On the other hand, $M_{s}^{\mathrm{Average'19}}$, obtained as a weighted
average from both lattice simulations and sum rule results, shows
better agreement with the experiment, and a reduction of the total
errors by about 40\%. This result for the mass difference sets a less
constraining bound

\begin{equation}
G_{bs}\lesssim\frac{1}{\left(220\,\mathrm{TeV}\right)^{2}}\,.
\end{equation}
The resulting constraints will be shown as blue regions over the parameter space.

\begin{figure}[!t]
\subfloat[\label{fig:theoreticallyclean_gbbgLmumu_plane_a}]{
\includegraphics[scale=0.45]{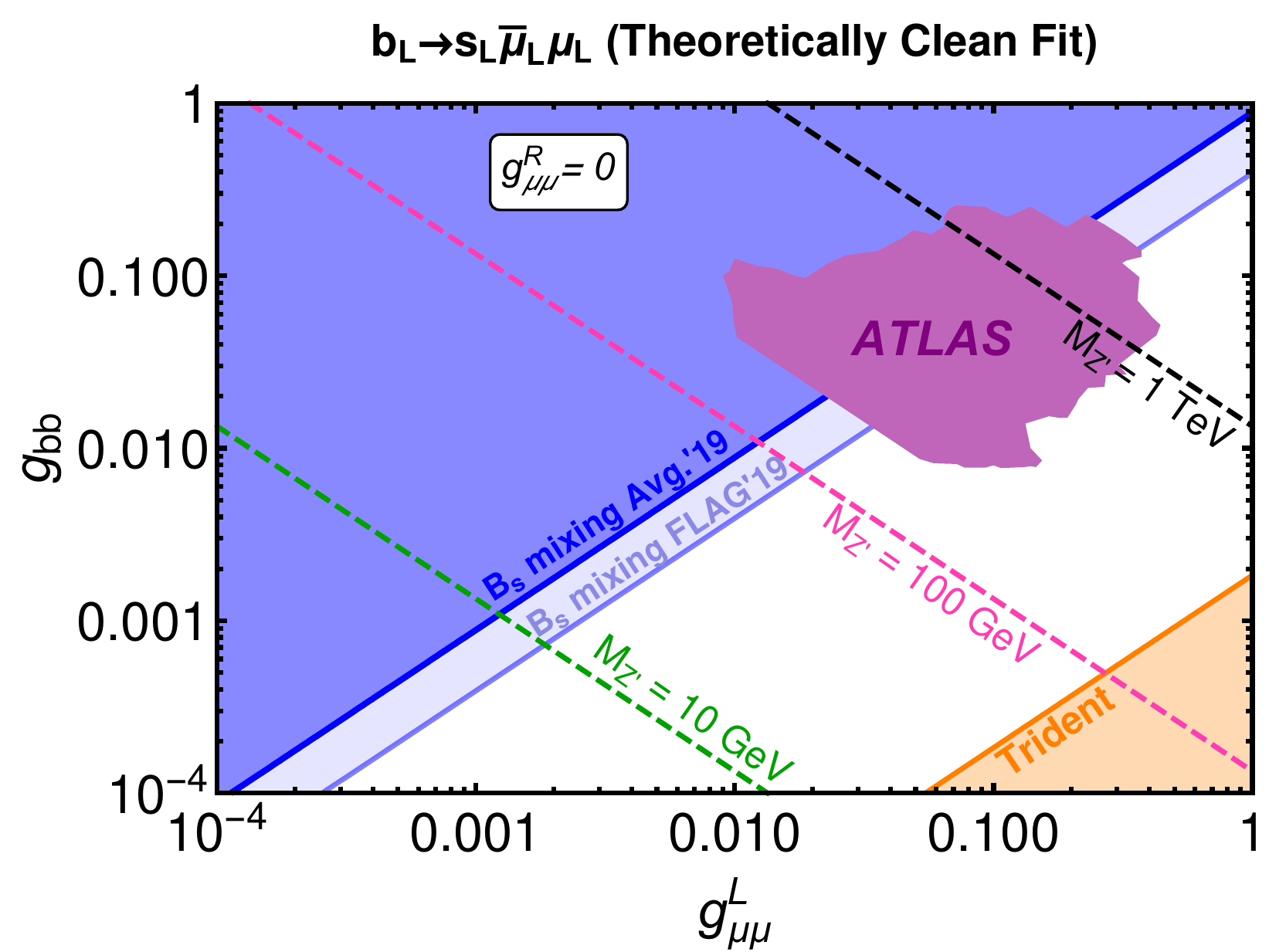}
}\subfloat[\label{fig:theoreticallyclean_gbbgLmumu_plane_b}]{\includegraphics[scale=0.45]{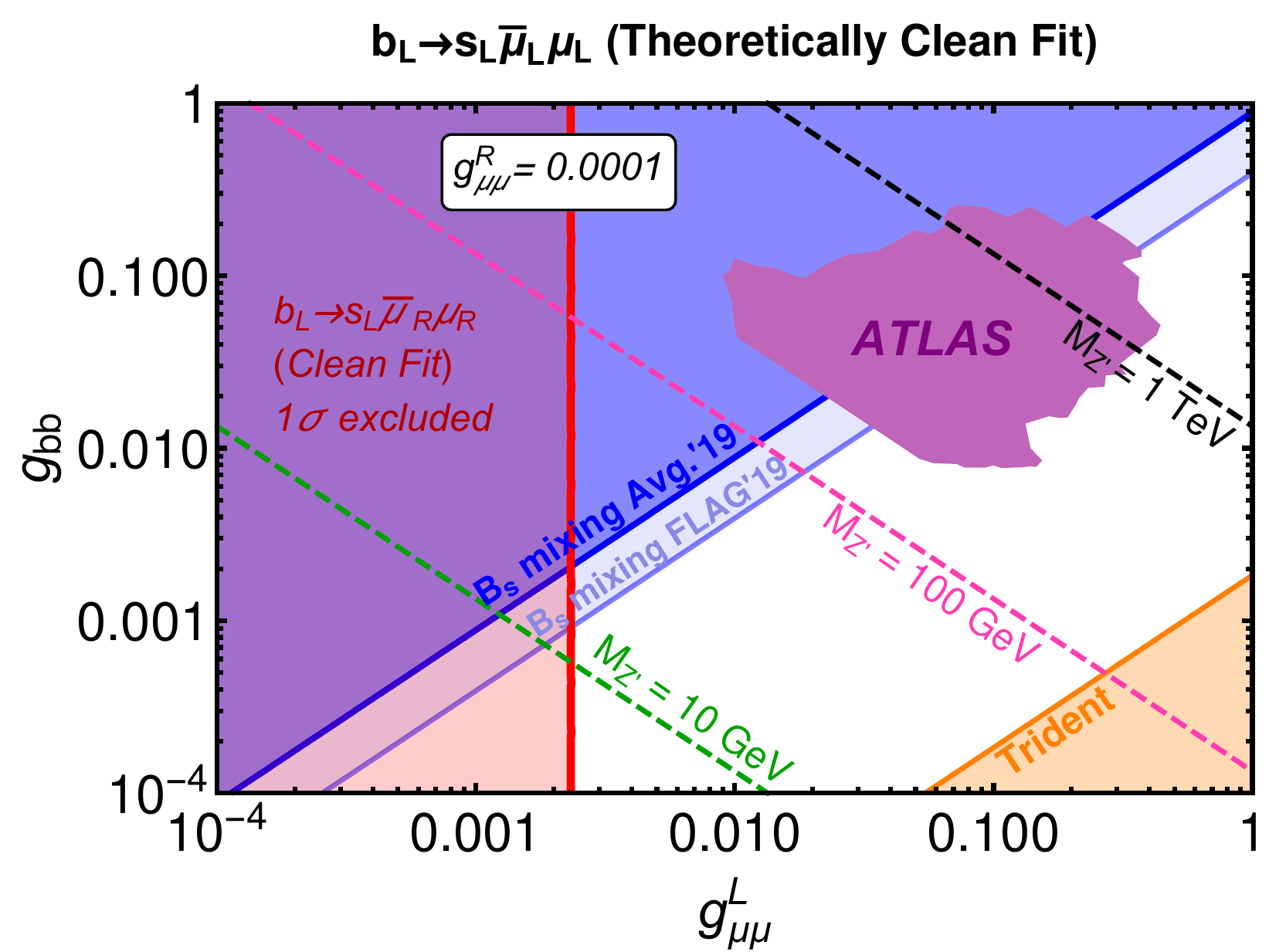}

}
\par
\subfloat[\label{fig:theoreticallyclean_gbbgLmumu_plane_c}]{
\includegraphics[scale=0.45]{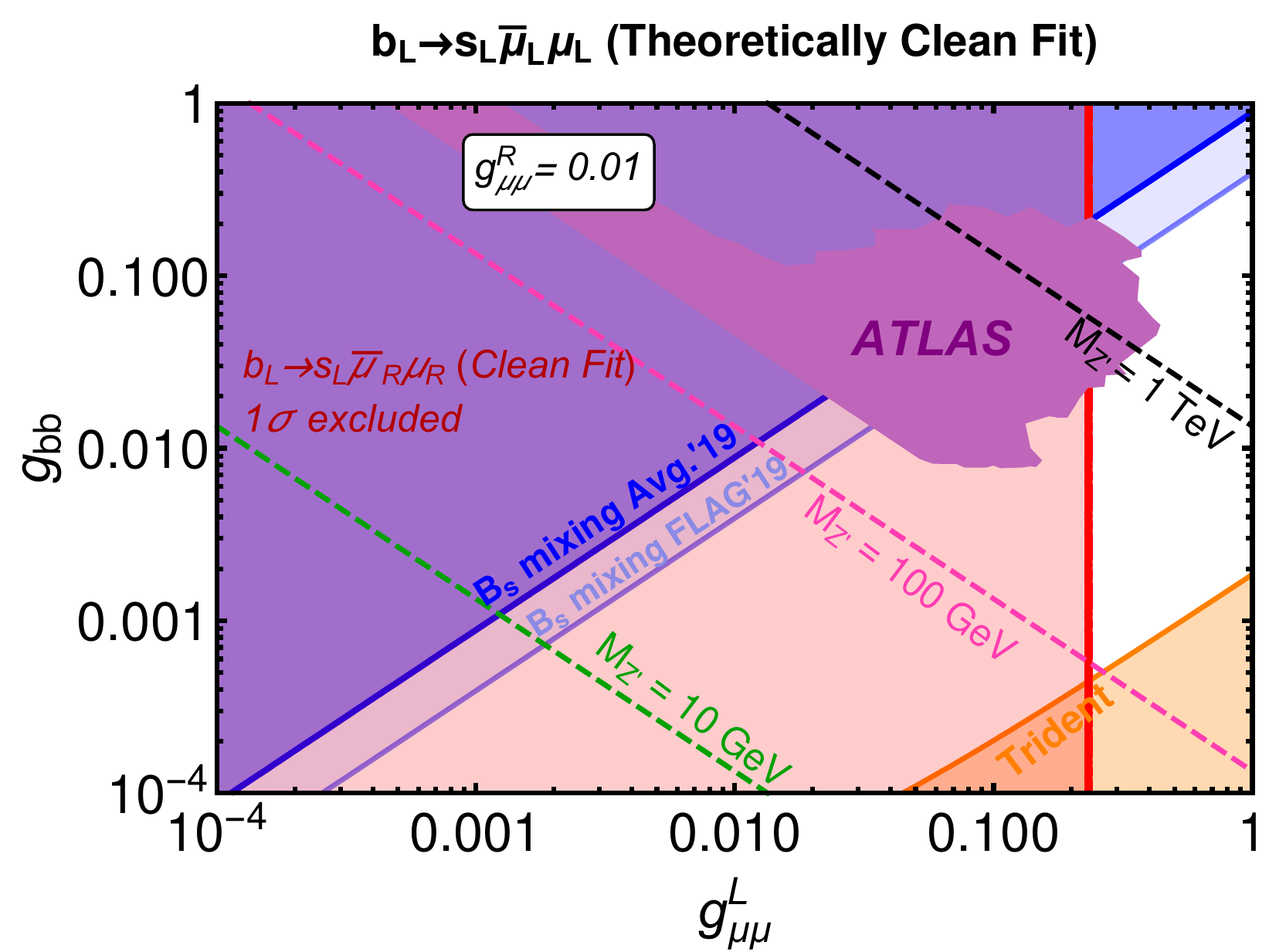}
}\subfloat[\label{fig:globalfit_gbbgLmumu_plane}]{\includegraphics[scale=0.45]{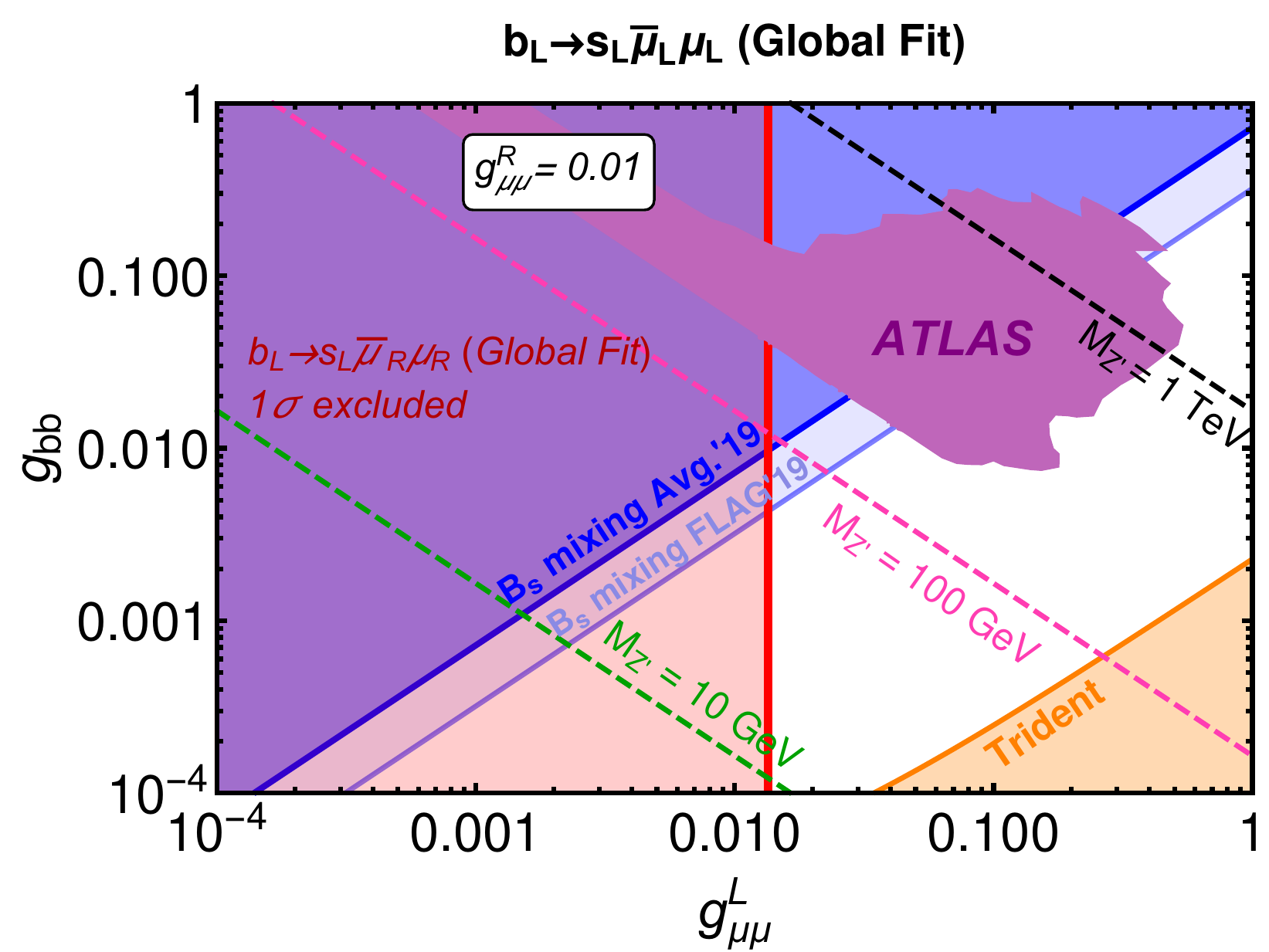}

}
\caption{The parameter space in the ($g_{\mu\mu}^{L},g_{bb}$) plane compatible
with $R_{K^{(*)}}$ anomalies and flavour constraints (white). The
$Z'$ mass varies over the plane, with an unique $Z'$ mass for each
point in the plane as required to match the best fit value for $G_{bs\mu}^{L}$(Eq.~\eqref{eq:GLbsmu})
of the theoretically clean fit in Table~\ref{tab:Theoretically-clean-fit}
(Fig.~\ref{fig:theoreticallyclean_gbbgLmumu_plane_a}, \ref{fig:theoreticallyclean_gbbgLmumu_plane_b},
\ref{fig:theoreticallyclean_gbbgLmumu_plane_c}) and the global fit
in Table~\ref{tab:Fit-including-angular} (Fig.~\ref{fig:globalfit_gbbgLmumu_plane}).
We show the recent $B_{s}-\bar{B}_{s}$ mixing constraints (blue and
light blue) \cite{DiLuzio:2019jyq}, the neutrino trident bounds (orange)
\cite{Falkowski:2017pss}, and the region excluded by LHC dimuon resonance
searches (purple)~\cite{ATLAS:2017fih}. When a non-vanishing $g_{\mu\mu}^{R}$
is considered, the red-shaded region is excluded of the $\text{1\ensuremath{\sigma}}$
range of $G_{bs\mu}^{R}$(Eq.~\eqref{eq:GRbsmu}) for the considered
fit. The dashed lines correspond to constant values of $M_{Z'}$ as
specified in the plots. \label{fig:gLmumu_gbb_plane}}
\end{figure}

\subsection{Neutrino trident}

The $Z'$ couplings to the second generation of the SM lepton doublet and singlet lead to a new
tree-level contribution (Fig.~\ref{fig: Z' exchange to trident})
to the effective 4-lepton interaction

\begin{equation}
\Delta\mathcal{L}_{\mathbf{\mathrm{eff}}}\supset-\frac{\left(g_{\mu\mu}^{L}\right)^{2}}{2M_{Z'}^{2}}\left(\bar{\mu}_{L}\gamma_{\mu}\mu_{L}\right)\left(\bar{\nu}_{\mu L}\gamma^{\mu}\nu_{\mu L}\right)-\frac{g_{\mu\mu}^{R}g_{\mu\mu}^{L}}{M_{Z'}^{2}}\left(\bar{\mu}_{R}\gamma_{\mu}\mu_{R}\right)\left(\bar{\nu}_{\mu L}\gamma^{\mu}\nu_{\mu L}\right)\,.
\end{equation}
This operator is constrained by the trident production $\nu_{\mu}\gamma^{*}\rightarrow\nu_{\mu}\mu^{+}\mu^{-}$
\cite{CHARM-II:1990dvf,CCFR:1991lpl,Altmannshofer:2014pba} . Using
the results of the global fit in Ref.~\cite{Falkowski:2017pss},
the bound over $g_{\mu\mu}^{L}$ and $g_{\mu\mu}^{R}$ is given by

\begin{equation}
-\frac{1}{\left(390\,\mathrm{GeV}\right)^{2}}\apprle\frac{\left(g_{\mu\mu}^{L}\right)^{2}+g_{\mu\mu}^{L}g_{\mu\mu}^{R}}{M_{Z'}^{2}}\apprle\frac{1}{\left(370\,\mathrm{GeV}\right)^{2}}\,,\label{eq:trident}
\end{equation}
whereas in our case only the right side of (\ref{eq:trident}) applies,
since according to Eqs.~(\ref{eq:gLmumu}) and (\ref{eq:gRmumu})
$g_{\mu\mu}^{L}$ and $g_{\mu\mu}^{R}$ have the same relative sign in our model
and hence the product $g_{\mu\mu}^{L}g_{\mu\mu}^{R}$ is positive.
The resulting constraints will be shown as orange regions over the parameter space.

\subsection{Constraints from lepton flavour violation}

Within the lepton sector the $Z'$ only couples to muons, hence no $Z'$ lepton flavour-violating couplings are generated. Therefore, in our $Z'$ model there are no contributions to lepton flavour-violating processes such as $\mu\rightarrow e\gamma$ or
$\tau\rightarrow3\mu$.

\subsection{Collider constraints}

Our model is not constrained by electron collider searches since our $Z'$
does not couple to electrons. However, further constraints on our
model come from LHC searches. For light $Z'$ masses, the LHC measurements
of the $Z$ decays to four muons, with the second muon pair produced
in the SM via a virtual photon \cite{CMS:2012bw,ATLAS:2014jlg}, $pp\rightarrow Z\rightarrow4\mu$,
set relevant constraints in the low mass region of $Z'$ models,
$5\lesssim M_{Z'}\apprle70\,\mathrm{GeV}$ \cite{Falkowski:2018dsl,Altmannshofer:2014cfa,Altmannshofer:2014pba,Altmannshofer:2016jzy}. We avoid such a constraint
by keeping $M_{Z'}>75\,\mathrm{GeV}$ in our analysis.\\

For heavier $Z'$ masses, the strongest constraints come from LHC
dimuon resonance searches, $pp\rightarrow Z'\rightarrow\mu^{+}\mu^{-}$,
see also \cite{Abdullah:2017oqj,Alonso:2017uky}. In our model, the $Z'$ is dominantly
produced at the LHC through its coupling to bottom quarks, $b\bar{b}\rightarrow Z'$.
The cross section $\sigma\left(b\bar{b}\rightarrow Z'\right)$ from
$b\bar{b}$ collisions is given for $g_{bb}=1$ in Fig. 3 of Ref.~\cite{Faroughy:2016osc},
we multiply it by $g_{bb}^{2}$ in order to obtain the cross section
for any $g_{bb}$. We neglect a further contribution coming from $b\bar{s}\rightarrow Z'$
since it is CKM suppressed by $V_{ts}^{2}$. Therefore, we assume that $\sigma\left(pp\rightarrow Z'\right)$
is dominated by the subprocess $b\bar{b}\rightarrow Z'$.
The $Z'$ boson can subsequently decay into muons, muon neutrinos,
bottom quarks, bottom-strange quark pair, and also into top quarks when kinematically
allowed. The partial decay widths are given by

\begin{align}
\begin{aligned}{} & \Gamma_{Z'\rightarrow\mu\bar{\mu}}=\frac{1}{24\pi}\left[\left(g_{\mu\mu}^{L}\right)^{2}+\left(g_{\mu\mu}^{R}\right)^{2}\right]M_{Z'}\,,\\
{} & \Gamma_{Z'\rightarrow\nu_{\mu}\bar{\nu}_{\mu}}=\frac{1}{24\pi}\left(g_{\mu\mu}^{L}\right)^{2}M_{Z'}\,,\\
{} & \Gamma_{Z'\rightarrow b\bar{b}}=\frac{1}{8\pi}g_{bb}^{2}M_{Z'}\,,\quad\Gamma_{Z'\rightarrow b\bar{s}}=\frac{1}{8\pi}g_{bb}^{2}V_{ts}^{2}M_{Z'}\,,\\
{} & \Gamma_{Z'\rightarrow t\bar{t}}=\frac{1}{8\pi}g_{bb}^{2}M_{Z'}\left(1-\frac{m_{t}^{2}}{M_{Z'}^{2}}\right)\sqrt{1-\frac{4m_{t}^{2}}{M_{Z'}^{2}}}\,,
\end{aligned}
\label{eq: full lagrangian-1}
\end{align}
from which we compute $\mathrm{Br}\left(Z'\rightarrow\mu\bar{\mu}\right)$
analytically,

\begin{equation}
\mathrm{Br}\left(Z'\rightarrow\mu\bar{\mu}\right)=\frac{\Gamma_{Z'\rightarrow\mu\bar{\mu}}}{\Gamma_{Z'\rightarrow\mu\bar{\mu}}+\Gamma_{Z'\rightarrow\nu_{\mu}\bar{\nu}_{\mu}}+\Gamma_{Z'\rightarrow b\bar{b}}+\Gamma_{Z'\rightarrow b\bar{s}}+\Gamma_{Z'\rightarrow t\bar{t}}}\,.
\end{equation}
Then $\sigma\left(pp\rightarrow Z'\rightarrow\mu^{+}\mu^{-}\right)$
is estimated using the narrow-width approximation,

\begin{equation}
\sigma\left(pp\rightarrow Z'\rightarrow\mu^{+}\mu^{-}\right)\approx\sigma\left(pp\rightarrow Z'\right)\mathrm{Br}\left(Z'\rightarrow\mu\bar{\mu}\right)\,,
\end{equation}
and compared with the limits obtained from the dimuon resonance search
by ATLAS \cite{ATLAS:2017fih}, which allows us to constrain $Z'$
masses between 150 GeV and 5 TeV. Previous studies \cite{Falkowski:2018dsl}
verified that the analogous Tevatron analyses give weaker constraints
than LHC. All things considered, the resulting ATLAS constraints will be shown
as purple regions over the parameter space.

\subsection{Higgs diphoton decay}

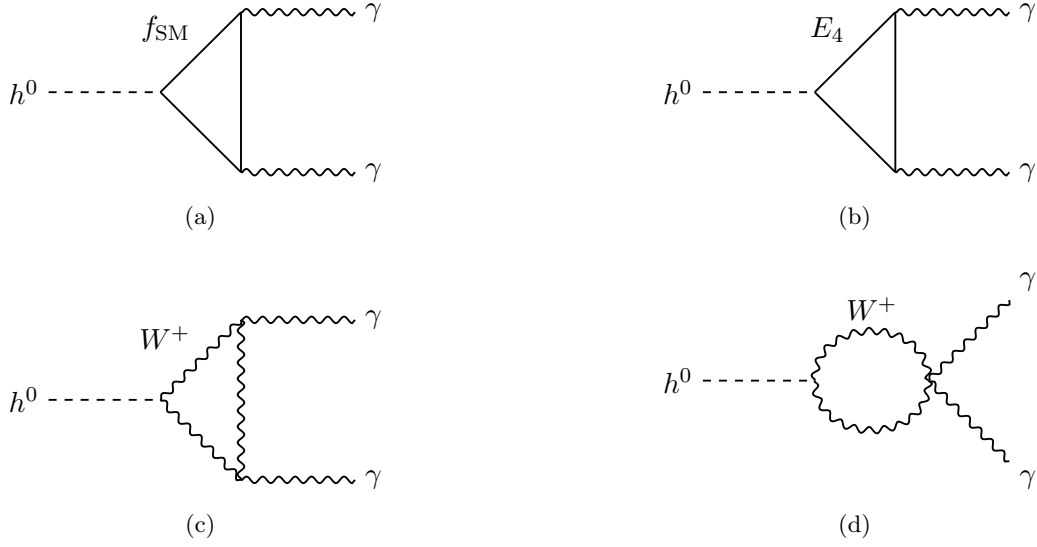
\begin{figure}[!t]
\subfloat[\label{fig:h2photons SMfermions}]{\begin{centering}
\begin{tikzpicture}
	\begin{feynman}
		\vertex (a) {\(h^{0}\)};
		\vertex [right=18mm of a] (b);
		\vertex [above right=of b] (f1);
		\vertex [below right=of b] (g1);
		\vertex [right= of f1] (f2) {\(\gamma\)};
		\vertex [right= of g1] (g2) {\(\gamma\)};
		\diagram* {
			(a) -- [scalar, line width=0.25mm] (b) -- [plain, edge label=\(f_{\mathrm{SM}}\), line width=0.25mm] (f1),
			(b) --  [plain, line width=0.25mm] (g1),
			(f1) -- [boson, line width=0.25mm] (f2),
			(g1) -- [boson, line width=0.25mm] (g2),
			(g1) -- [plain, line width=0.25mm] (f1),
	};
	\end{feynman}
\end{tikzpicture}
\par\end{centering}
}$\qquad$$\qquad\qquad\qquad$\subfloat[\label{fig:h2photons VLfermion}]{\begin{centering}
\begin{tikzpicture}
	\begin{feynman}
		\vertex (a) {\(h^{0}\)};
		\vertex [right=18mm of a] (b);
		\vertex [above right=of b] (f1);
		\vertex [below right=of b] (g1);
		\vertex [right= of f1] (f2) {\(\gamma\)};
		\vertex [right= of g1] (g2) {\(\gamma\)};
		\diagram* {
			(a) -- [scalar, line width=0.25mm] (b) -- [plain, edge label=\(E_{4}\), line width=0.25mm] (f1),
			(b) --  [plain, line width=0.25mm] (g1),
			(f1) -- [boson, line width=0.25mm] (f2),
			(g1) -- [boson, line width=0.25mm] (g2),
			(g1) -- [plain, line width=0.25mm] (f1),
	};
	\end{feynman}
\end{tikzpicture}
\par\end{centering}
}

\subfloat[\label{fig:h2photons Wa}]{\begin{centering}
\begin{tikzpicture}
	\begin{feynman}
		\vertex (a) {\(h^{0}\)};
		\vertex [right=18mm of a] (b);
		\vertex [above right=of b] (f1);
		\vertex [below right=of b] (g1);
		\vertex [right= of f1] (f2) {\(\gamma\)};
		\vertex [right= of g1] (g2) {\(\gamma\)};
		\diagram* {
			(a) -- [scalar, line width=0.25mm] (b) -- [boson, edge label=\(W^{+}\), line width=0.25mm] (f1),
			(b) --  [boson, line width=0.25mm] (g1),
			(f1) -- [boson, line width=0.25mm] (f2),
			(g1) -- [boson, line width=0.25mm] (g2),
			(g1) -- [boson, line width=0.25mm] (f1),
	};
	\end{feynman}
\end{tikzpicture}
\par\end{centering}
}$\qquad\qquad\qquad\qquad$\subfloat[\label{fig:h2photons Wb}]{\begin{centering}
\begin{tikzpicture}
	\begin{feynman}
		\vertex (a) {\(h^{0}\)};
		\vertex [right=18mm of a] (b);
		\vertex [right=of b] (c);
		\vertex [above right=of c] (f1) {\(\gamma\)};
		\vertex [below right=of c] (g1) {\(\gamma\)};
		\diagram* {
			(a) -- [scalar, line width=0.25mm] (b) -- [boson, half left, edge label=\(W^{+}\), line width=0.25mm] (c),
			(b) --  [boson, half right, line width=0.25mm] (c),
			(c) -- [boson, line width=0.25mm] (f1),
			(c) -- [boson, line width=0.25mm] (g1),
	};
	\end{feynman}
\end{tikzpicture}
\par\end{centering}
}\caption{Diagrams contributing to the Higgs diphoton decay ($h^{0}\rightarrow\gamma\gamma$)
where $f_{\mathrm{SM}}=u_{i},d_{i},e_{i}$, $i=1,2,3$ and $E_{4}$
is the 4th family VL lepton. \label{fig: h2photons}}
\end{figure}

After Spontaneous Symmetry Breaking (SSB), the Yukawa term in Eq.~(\ref{eq: full lagrangian}) involving
the SM Higgs field and the 4th VL lepton gives rise to the chirality-flipping
mass $M_{4}^{C}$, which gives a very important contribution in Eq.~(\ref{eq:g-2})
for accommodating $\Delta a_{\mu}$ with the experimental measurements.
On the other hand, $M_{4}^{C}$ is also expected to give an extra
contribution to the decay of the Standard Model Higgs to two photons,
a process that has been explored in colliders. Firstly, within the
SM, fermions (Fig.~\ref{fig:h2photons SMfermions}) and $W^{\pm}$
bosons (Figs.~\ref{fig:h2photons Wa}, \ref{fig:h2photons Wb}) contribute
to the decay channel $h^{0}\rightarrow\gamma\gamma$ \cite{Gunion:1989we}

\begin{equation}
\Gamma(h^{0}\rightarrow\gamma\gamma)_{\mathrm{SM}}=\frac{\alpha^{2}m_{h}^{3}}{256\pi^{3}v^{2}}\left|F_{1}(\tau_{W})+\sum_{f\,\epsilon\,\mathrm{SM}}N_{cf}Q_{f}^{2}F_{1/2}(\tau_{f})\right|^{2}\,,
\end{equation}
where $\alpha=1/137$, $m_{h}=126\,\mathrm{GeV}$, $v=246\,\mathrm{GeV}$,
$N_{cf}=1\text{\,(leptons)},\,3\text{ (quarks)}$, $Q_{f}$ is the electromagnetic
charge of the fermion $f$ in units of $e$, and the loop functions are defined as

\begin{equation}
F_{1}=2+3\tau+3\tau\left(2-\tau\right)f(\tau)\,,
\end{equation}

\begin{equation}
F_{1/2}=-2\tau\left[1+\left(1-\tau\right)f(\tau)\right]\,,
\end{equation}
with 

\begin{equation}
\tau_{i}=4m_{i}^{2}/m_{h}^{2}
\end{equation}
\\ 

and

\begin{equation}
f(\tau)=\left\{ \begin{array}{c}
{\displaystyle \left[\arcsin\left(1/\sqrt{\tau}\right)\right]^{2}\,,\quad\text{if }\quad\tau\geq1\,,}\\
\,\\
{\displaystyle -\frac{1}{4}\left[\ln\left(\frac{1+\sqrt{1-\tau}}{1-\sqrt{1-\tau}}\right)-i\pi\right]^{2}\,,\quad\text{if }\quad\tau<1\,.}
\end{array}\right.
\end{equation}
Note here that for large $\tau$, $F_{1/2}\rightarrow-4/3$. The dominant
contribution to $\Gamma(h^{0}\rightarrow\gamma\gamma)_{\mathrm{SM}}$
is the contribution of the $W$ bosons,

\begin{equation}
F_{1}(\tau_{W})\simeq8.33\,,
\end{equation}
and it interferes destructively with the top-quark loop

\begin{equation}
N_{ct}Q_{t}^{2}F_{1/2}(\tau_{f})=3\left(2/3\right)^{2}\left(-1.37644\right)=-1.83526\,,
\end{equation}
therefore

\begin{equation}
\Gamma(h^{0}\rightarrow\gamma\gamma)_{\mathrm{SM}}=\frac{\alpha^{2}m_{h}^{3}}{256\pi^{3}v^{2}}\left|8.33-1.83526\right|^{2}\simeq9.15636\cdot10^{-6}\,\mathrm{GeV}\,.
\end{equation}
The exact result by taking into account the contribution of all SM
fermions is

\begin{equation}
\Gamma(h^{0}\rightarrow\gamma\gamma)_{\mathrm{SM}}=9.34862\cdot10^{-6}\,\mathrm{GeV}\,,
\end{equation}
and if we take $\Gamma(h^{0}\rightarrow\mathrm{all})_{\mathrm{PDG\,2021}}=3.2_{-2.2}^{+2.8}\,\mathrm{MeV}$,
then

\begin{equation}
\mathrm{BR}\left(h^{0}\rightarrow\gamma\gamma\right)_{\mathrm{SM}}=\frac{\Gamma(h^{0}\rightarrow\gamma\gamma)_{\mathrm{SM}}}{\Gamma(h^{0}\rightarrow\mathrm{all})_{\mathrm{PDG\,2021}}}\times100\simeq0.29\,\%\,.
\end{equation}

Now we add the contribution of a fourth VL lepton (Fig.~\ref{fig:h2photons VLfermion})
with VL mass $M_{4}^{L}$ that couples to the Higgs
via the chirality-flipping mass $M_{4}^{C}$, where $M_{4}^{L}\gg M_{4}^{C}$
(in such a way that the propagating mass of the fourth lepton can be approximated
by the VL mass)~\cite{Bizot:2015zaa},

\begin{equation}
\Gamma(h^{0}\rightarrow\gamma\gamma)=\frac{\alpha^{2}m_{h}^{3}}{256\pi^{3}v^{2}}\left|F_{1}(\tau_{W})+\sum_{f\,\epsilon\,\mathrm{SM}}N_{cf}Q_{f}^{2}F_{1/2}(\tau_{f})+\frac{M_{4}^{C}}{M_{4}^{L}}F_{1/2}\left(\tau_{E_{4}}\right)\right|^{2}\,.
\end{equation}
We can see that the new contribution proportional to the chirality-flipping
mass is suppressed by the heavier VL mass. Moreover, this new contribution
decreases ${\Gamma(h^{0}\rightarrow\gamma\gamma)}$, since it interferes
destructively with the most sizable contribution of the $W$ bosons.
Let us now compare with the experimental results for the $h^{0}$
signal strength in the $h^{0}\rightarrow\gamma\gamma$ channel \cite{ParticleDataGroup:2020ssz},

\begin{equation}
R_{\gamma\gamma}=\frac{\Gamma(h^{0}\rightarrow\gamma\gamma)}{\Gamma(h^{0}\rightarrow\gamma\gamma)_{\mathrm{SM}}}\,,
\end{equation}

\begin{equation}
R_{\gamma\gamma}^{\mathrm{PDG,\,2020}}=1.11{}_{-0.09}^{+0.1}\,.
\end{equation}
In the case of $M_{4}^{C}=200\,\mathrm{GeV}$,

\begin{equation}
R_{\gamma\gamma}^{\mathrm{VL}}=\frac{\left|8.33-1.83526-0.0533333\right|^{2}}{\left|8.33-1.83526\right|^{2}}=0.983813\,.
\end{equation}
In the case of $M_{4}^{C}=600\,\mathrm{GeV}$,

\begin{equation}
R_{\gamma\gamma}^{\mathrm{VL}}=\frac{\left|8.33-1.83526-0.16\right|^{2}}{\left|8.33-1.83526\right|^{2}}=0.951336\,.
\end{equation}
Therefore, even for a value of $M_{4}^{C}$ close to the perturbation
theory limit $M_{4}^{C}\lesssim\sqrt{4\pi}v/\sqrt{2}\simeq616.8\,\mathrm{GeV}$,
the chirality-flipping mass contribution to $h^{0}\rightarrow\gamma\gamma$
is within the $2\sigma$ range of $R_{\gamma\gamma}^{\mathrm{PDG,\,2020}}$.

\section{Results and discussion\label{sec:Results-and-discussion}}

\begin{figure}[!t]
\subfloat[]{\includegraphics[scale=0.44]{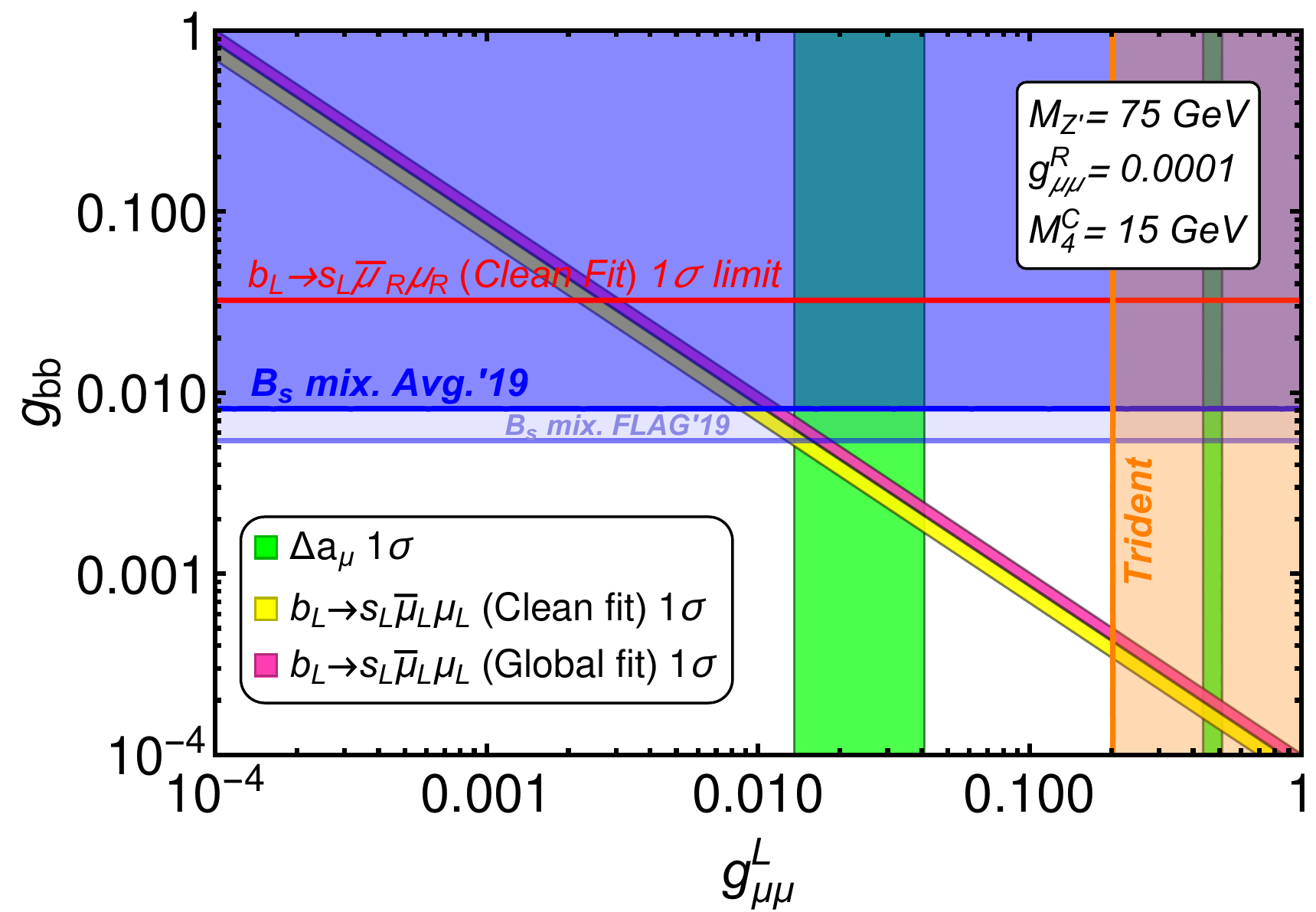}

}\subfloat[]{\includegraphics[scale=0.44]{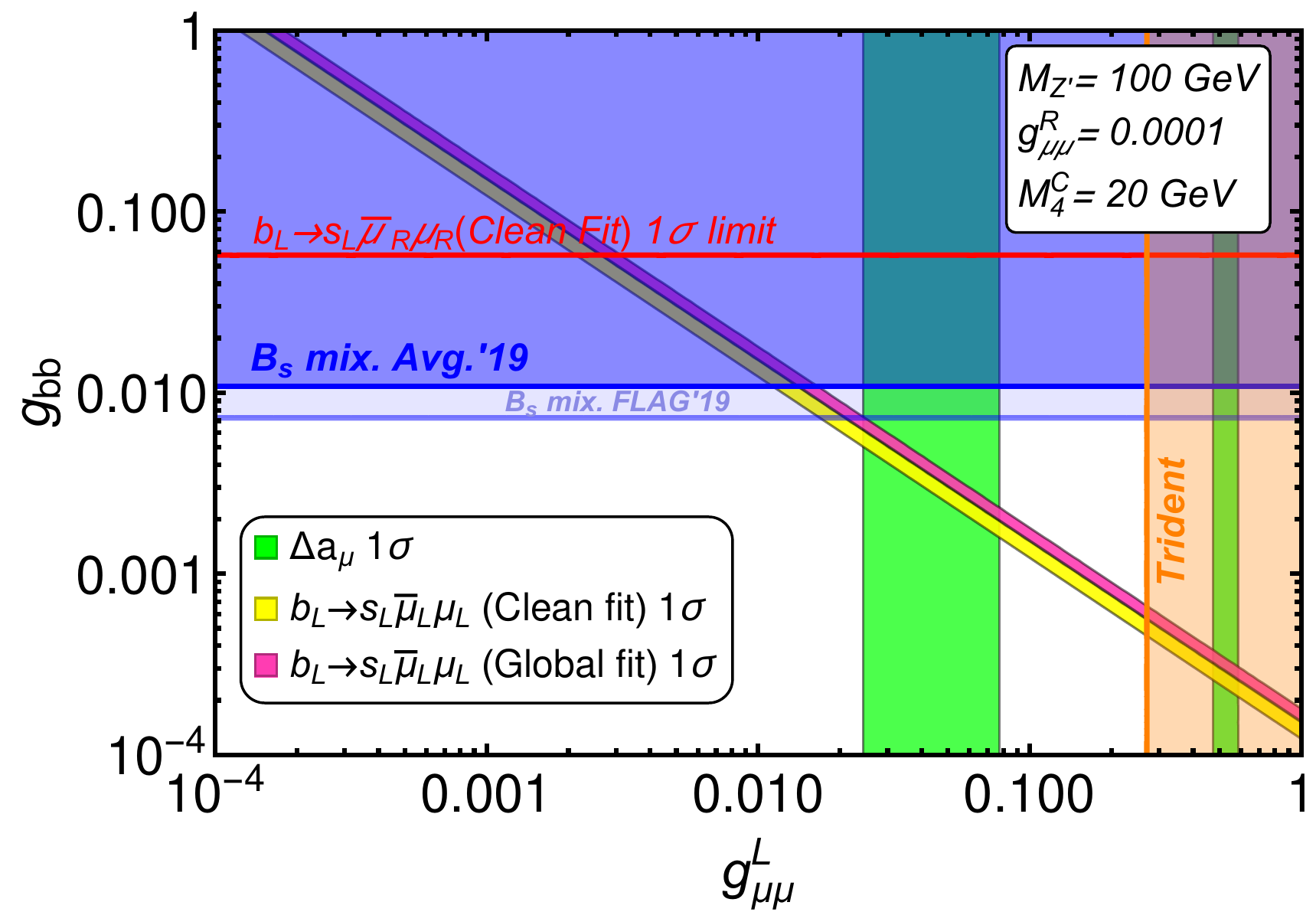}

}

\subfloat[]{\includegraphics[scale=0.44]{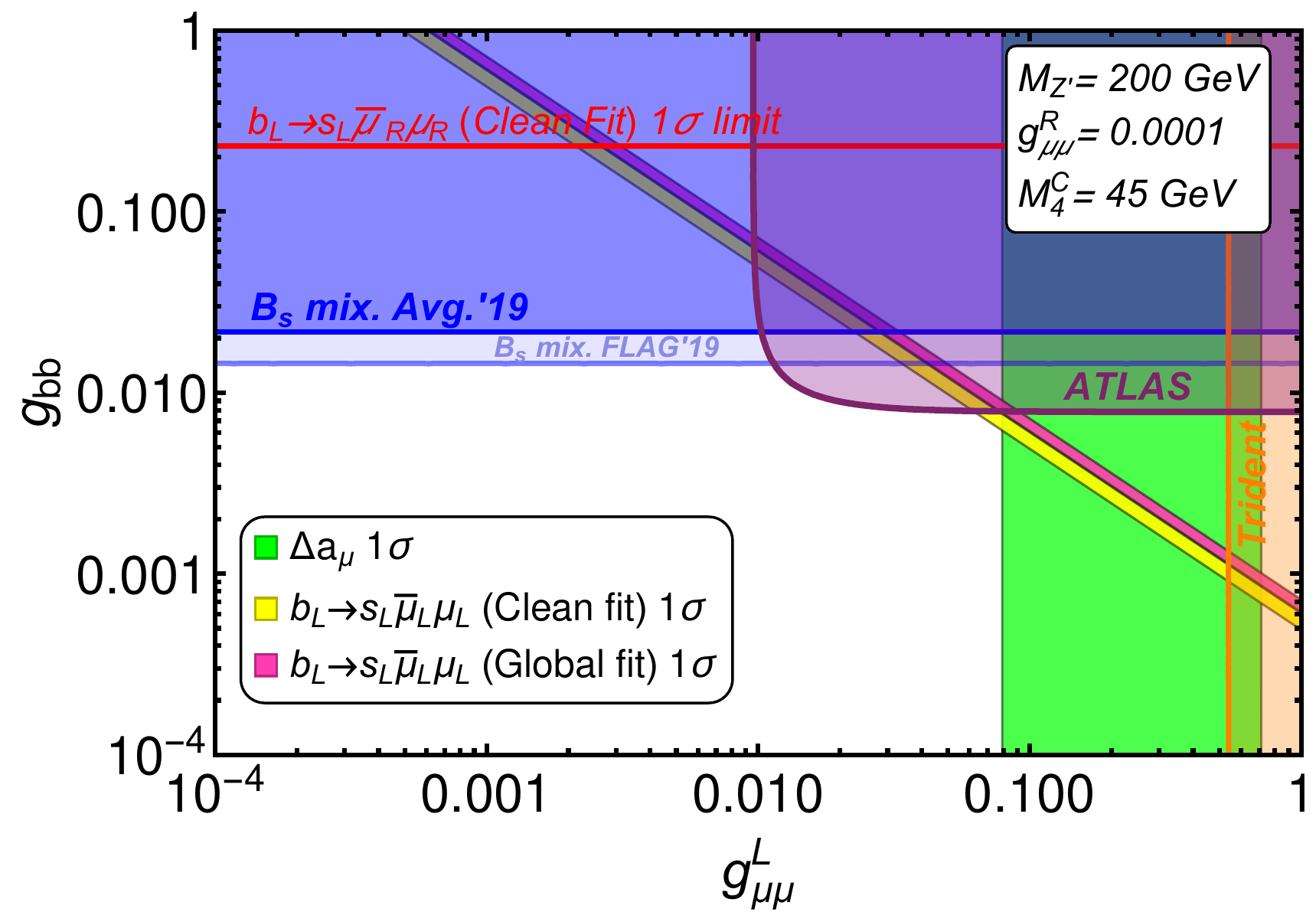}

}\subfloat[]{\includegraphics[scale=0.44]{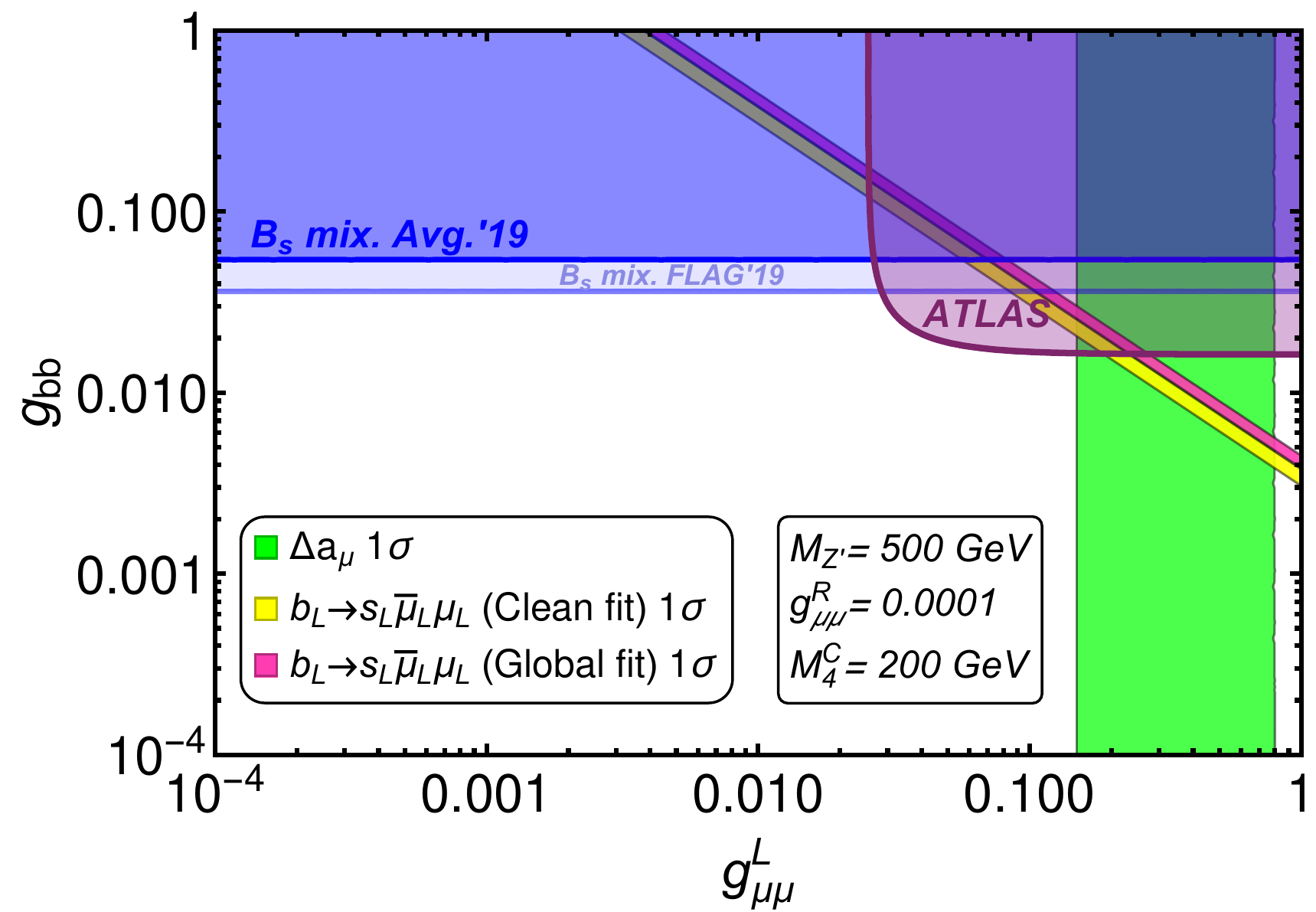}

}

\caption{Bounds on the parameter space in the ($g_{\mu\mu}^{L}$, $g_{bb}$)
plane for fixed $Z'$ masses: 75, 100, 200 and 500 GeV, as indicated
on each panel. Each panel also displays the considered $M_{4}^{C}$
and $g_{\mu\mu}^{R}$, while the propagating mass of the VL lepton
is always kept as ${m_{E}\simeq M_{4}^{L}=5\,\mathrm{TeV}}$. The green
region explains $\Delta a_{\mu}$ up to $1\sigma$. The yellow and
pink regions fit the Wilson coefficient $G_{bs\mu}^{L}$ (\ref{eq:GLbsmu})
up to $1\sigma$ for the theoretically clean fit and the global fit
\cite{Geng:2021nhg}, respectively. The red horizontal line shows
the limit of the $1\sigma$ region for the Wilson coefficient $G_{bs\mu}^{R}$
(\ref{eq:GRbsmu}) in the more restrictive theoretically clean fit,
in such a way that the parameter space above the red line is excluded.
The blue and orange areas show the $B_{s}-\bar{B}_{s}$ mixing \cite{DiLuzio:2019jyq}
and neutrino trident \cite{Falkowski:2017pss} exclusions, respectively,
while the purple region is excluded by LHC dimuon resonance searches  \cite{ATLAS:2017fih}.
\label{fig:Fitting_both_anomalies_LowMasses}}
\end{figure}

In Fig.~\ref{fig:gLmumu_gbb_plane} we have displayed the parameter
space in the $(g_{\mu\mu}^{L},\,g_{bb})$ plane for $g_{\mu\mu}^{R}=0,\,0.0001,$ $0.01$,
considering the theoretically clean fit (Figs.~\ref{fig:theoreticallyclean_gbbgLmumu_plane_a},
\ref{fig:theoreticallyclean_gbbgLmumu_plane_b}, \ref{fig:theoreticallyclean_gbbgLmumu_plane_c})
and the global fit (Fig.~\ref{fig:globalfit_gbbgLmumu_plane}). In
every case, there is parameter space free from all the constraints that is able to explain $R_{K^{(*)}}$. If we set $g_{\mu\mu}^{R}=0$
(Fig.~\ref{fig:theoreticallyclean_gbbgLmumu_plane_a}), we are making
a purely left-handed explanation of~$R_{K^{(*)}}$, hence recovering the same
results as in Ref.~\cite{Falkowski:2018dsl}. As $g_{\mu\mu}^{R}$
is increased, the condition of keeping the contribution to $b_{L}\rightarrow s_{L}\bar{\mu}_{R}\mu_{R}$
(namely the Wilson coefficient $G_{bs\mu}^{R}$) within the $1\sigma$
range becomes constraining over the parameter space, especially when
the theoretically clean fit is considered (Figs.~\ref{fig:theoreticallyclean_gbbgLmumu_plane_b},
\ref{fig:theoreticallyclean_gbbgLmumu_plane_c}). On the other hand,
if the global fit is considered (Fig.~\ref{fig:globalfit_gbbgLmumu_plane}),
then larger values of $g_{\mu\mu}^{R}$ are accessible.\\

\begin{figure}[!th]
\subfloat[]{\includegraphics[scale=0.44]{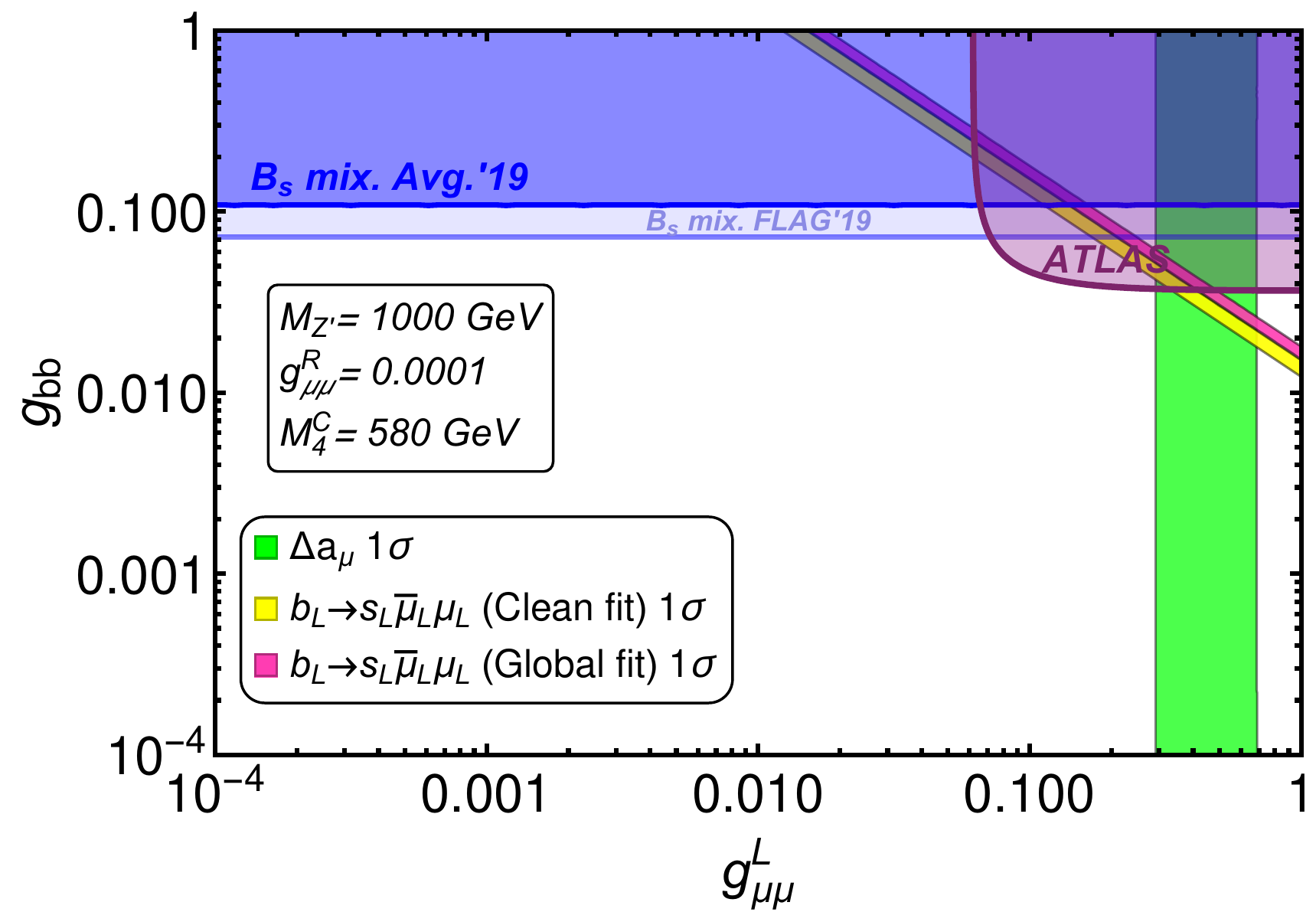}

}\subfloat[]{\includegraphics[scale=0.44]{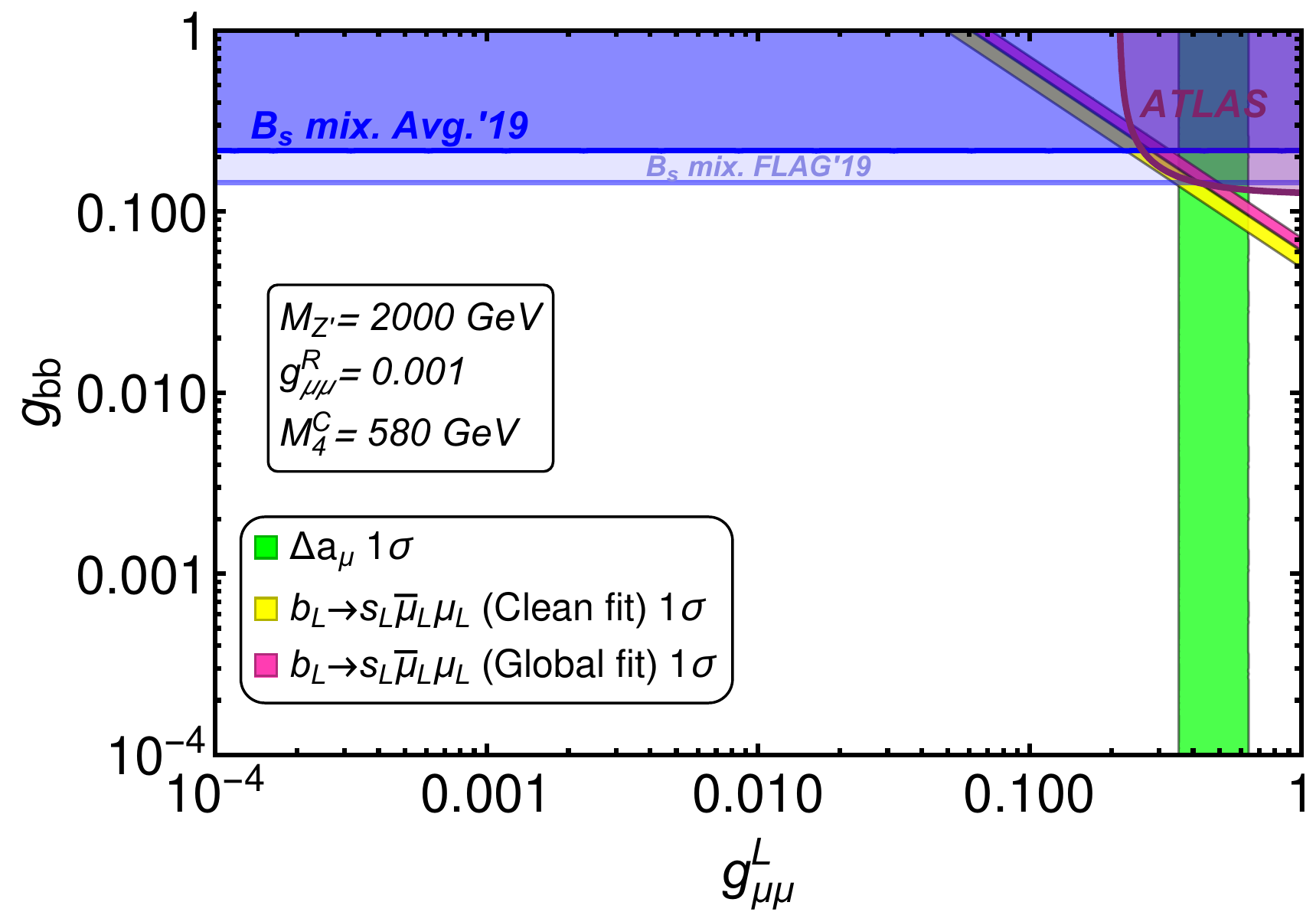}

}

\subfloat[\label{fig:theoreticallyclean_gbbMZ_plane-1-1}]{\includegraphics[scale=0.44]{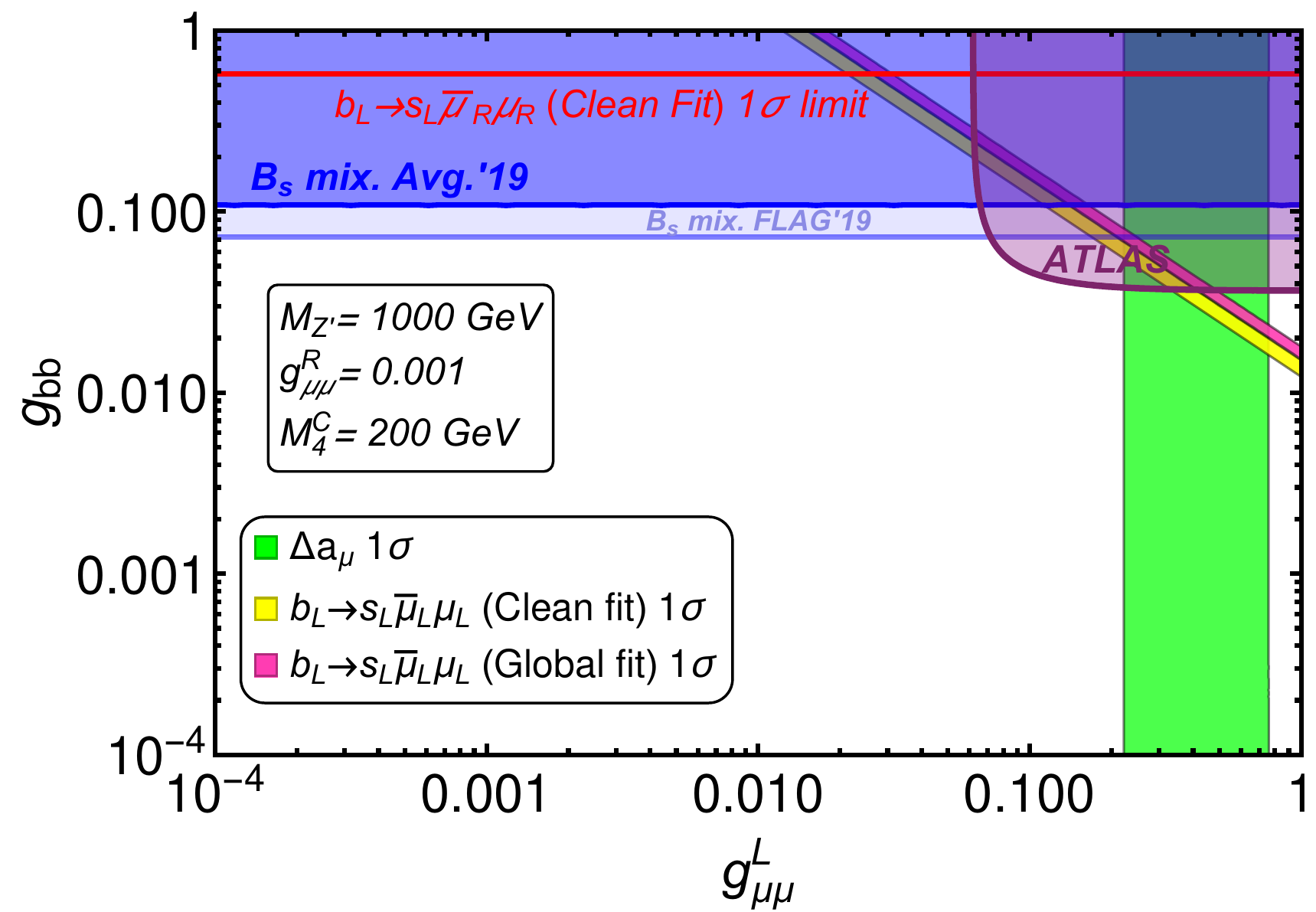} 

}\subfloat[\label{fig:globalfit_gbbgLMZ_plane-1-1}]{\includegraphics[scale=0.44]{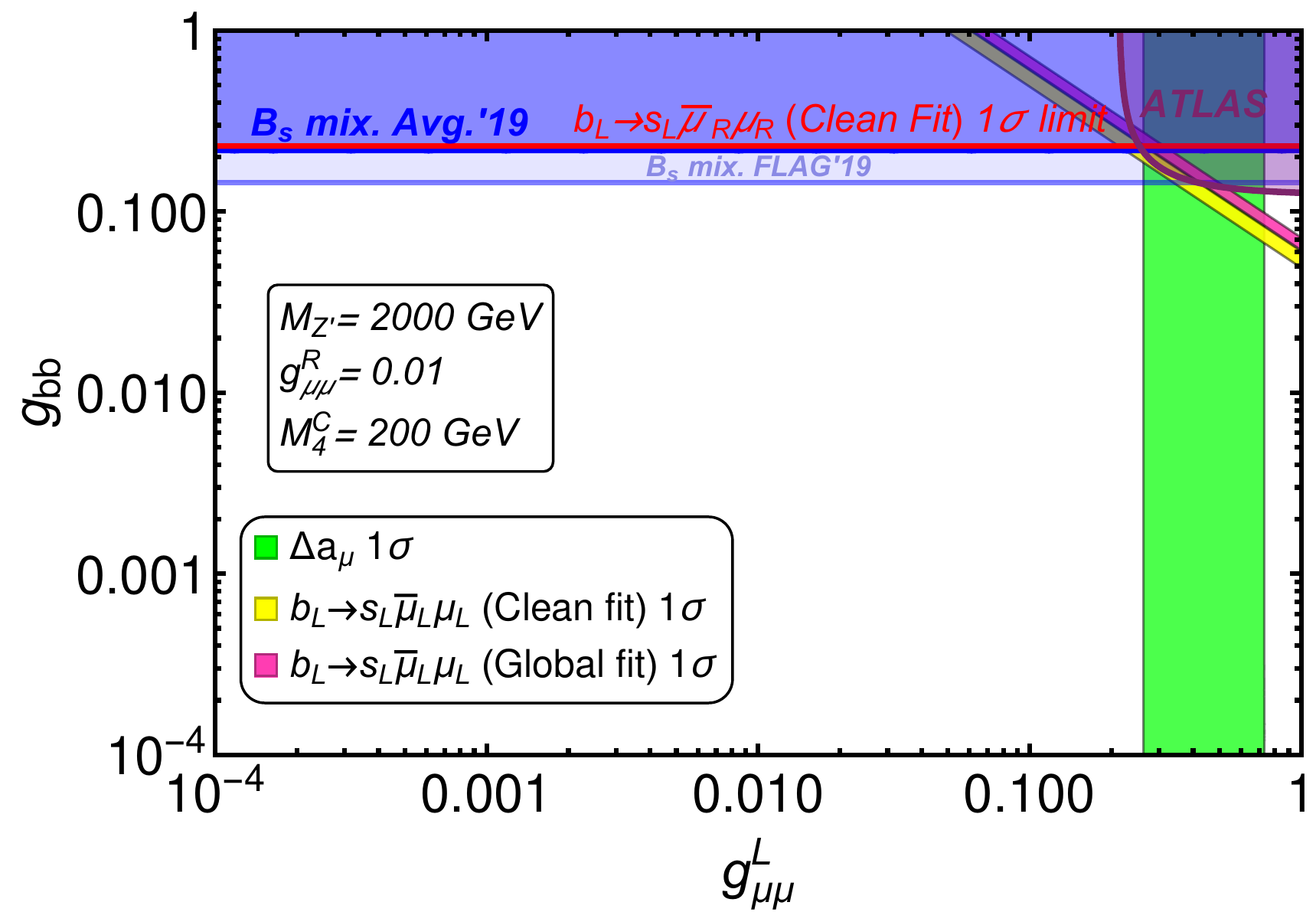}

}

\caption{The same as in Fig.~\ref{fig:Fitting_both_anomalies_LowMasses} but for heavy $Z'$ masses: 1000 and 2000 GeV, as indicated
on each panel.\label{fig:Fitting_both_anomalies_HeavyMasses}}
\end{figure}

In Figs.~\ref{fig:Fitting_both_anomalies_LowMasses} and \ref{fig:Fitting_both_anomalies_HeavyMasses}, for light and heavy $Z'$ masses respectively,
it can be seen that both the contribution to $b_{L}\rightarrow s_{L}\bar{\mu}_{L}\mu_{L}$
that explains $R_{K^{(*)}}$ (namely the Wilson Coefficient $G_{bs\mu}^{L}$)
and $\Delta a_{\mu}$ can be produced simultaneously within their $1\sigma$
region by a $Z'$ with a mass in the range of 75 GeV to 2 TeV, for
both the theoretically clean fit and the global fit. Within this range
of masses, $G_{bs\mu}^{L}$, $G_{bs\mu}^{R}$ and $\Delta a_{\mu}$
can be simultaneously fitted with the same parameters
up to the $1\sigma$ ranges of both the theoretically clean fit and
the global fit, since in all the considered cases the parameter space
where $G_{bs\mu}^{L}$ and $\Delta a_{\mu}$ are simultaneously explained
is also within the $1\sigma$ range of $G_{bs\mu}^{R}.$ The upper
bound for the latter in the theoretically clean fit is displayed in
Figs.~\ref{fig:Fitting_both_anomalies_LowMasses} and \ref{fig:Fitting_both_anomalies_HeavyMasses}
as a red horizontal line, there is no lower bound displayed since
$G_{bs\mu}^{R}$ is compatible with zero in both fits. In all the
explored cases, the condition of fitting $G_{bs\mu}^{R}$ is less
constraining than $B_{s}-\bar{B}_{s}$ mixing.\\
 
For light $Z'$ masses around $75-200$ GeV (Fig.~\ref{fig:Fitting_both_anomalies_LowMasses}),
both anomalies $R_{K^{(*)}}$ and $\Delta a_{\mu}$ can be
explained simultaneously with the condition of small $g_{\mu\mu}^{R}$
and $M_{4}^{C}$. On the other hand, as displayed in Fig.~\ref{fig:Fitting_both_anomalies_HeavyMasses},
for heavy $Z'$ masses of $1-2$ TeV both anomalies can also be explained simultaneously
but it is required to increase either $g_{\mu\mu}^{R}$ (two lower
panels of Fig.~\ref{fig:Fitting_both_anomalies_HeavyMasses}) or
$M_{4}^{C}$ (two upper panels of Fig.~\ref{fig:Fitting_both_anomalies_HeavyMasses}).
In every case, $M_{4}^{C}$ is kept below the perturbation theory limit
of $M_{4}^{C}\apprle\sqrt{4\pi}v/\sqrt{2}\approx618\,\mathrm{GeV}$,
and its contribution to Higgs diphoton decay has been previously proven
to be within the $2\sigma$ range of the experimental signals even
for values of $M_{4}^{C}$ close to the perturbation theory limit.
Moreover, in every case $g_{\mu\mu}^{R}$ is set to values for which $G_{bs\mu}^{R}$
can be simultaneously fitted, hence $R_{K^{(*)}}$ is explained. Fitting
both anomalies simultaneously for $M_{Z'}>2\,\mathrm{TeV}$ could
in principle be possible but would require chirality-flipping masses
too close to the perturbation theory limit, and/or values of
$g_{\mu\mu}^{R}$ of $\mathcal{O}(0.1)$ or higher, for which the
$\text{1\ensuremath{\sigma}}$ range of $G_{bs\mu}^{R}$ in the theoretically
clean fit becomes more challenging to fit, leading to constraints
over the parameter space larger than the present limits of $B_{s}-\bar{B}_{s}$
mixing. Instead, if we consider the global fit that includes angular
observables of $B\rightarrow K^{*}\bar{\mu}\mu$ data, here $G_{bs\mu}^{R}$ is compatible
with larger positive values and hence also larger $g_{\mu\mu}^{R}$
are accessible, in such a way that explaining both anomalies with heavier
masses of $Z'$ is possible. However, Figs.~\ref{fig:Fitting_both_anomalies_LowMasses}
and \ref{fig:Fitting_both_anomalies_HeavyMasses} also show that collider
constraints coming from dimuon resonance searches by ATLAS \cite{ATLAS:2017fih}
are very constraining for $M_{Z'}>500\,\mathrm{GeV}$. Although in
every case we can still find good points that simultaneously explain both $R_{K^{(*)}}$ and
$\Delta a_{\mu}$ while avoiding the ATLAS constraint, such
points could be ruled out in the future by the upcoming LHC run 3
starting in 2022.

\section{Conclusions\label{sec:Conclusions}}

We have shown that both muon anomalies $R_{K^{(*)}}$
and $\left(g-2\right)_{\mu}$ can be simultaneously addressed in a simplified 
fermiophobic $Z'$ model with $75\,\mathrm{GeV}\apprle M_{Z'}\apprle2\,\mathrm{TeV}$.
The explanation of $\left(g-2\right)_{\mu}$ in this model requires
non-vanishing couplings of $Z'$ to left-handed and right-handed muons
$g_{\mu\mu}^{L}$, $g_{\mu\mu}^{R}$, along with a non-vanishing chirality-flipping
mass $M_{4}^{C}$ obtained from the coupling of a fourth vector-like
lepton to the SM Higgs. The explanation of $R_{K^{(*)}}$ also requires
a coupling of $Z'$ to $bs$-quarks. Such $Z'$ couplings
are obtained in this model through mixing of muons and bottom quarks
with a fourth vector-like fermion family. 
In particular, the $Z'$ coupling to $bs$ is
CKM suppressed since $g_{bb}V_{ts}$ in the basis in which the up-quark mass matrix is diagonal. 

The scenario considered in this paper represents
a minimal mixing framework in which only three mixing parameters are
involved, By contrast, other $Z'$ models that address both muon anomalies are either
not fermiophobic \cite{Allanach:2015gkd,Raby:2017igl}, consider extra
symmetries \cite{CarcamoHernandez:2019xkb} or involve a general
mixing framework with a very large number of parameters \cite{Kawamura:2019hxp,Kawamura:2019rth},
where only a search of best fit points is performed. Instead, within the simplified 
approach followed here, we had been able to systematically explore the parameter space, extracting interesting conclusions in the process.

The fact that the explanation of the muon $g-2$ anomaly requires a non-zero coupling $g_{\mu\mu}^{R}\neq0$ means that it is not possible to provide a purely left-handed explanation of $R_{K^{(*)}}$ as in previous
studies \cite{King:2018fcg,Falkowski:2018dsl} which do not consider the muon $g-2$ anomaly. Consequently it is necessary here to 
fit both the LH and RH Wilson coefficients of the effective operators,
$G_{bs\mu}^{L}$ and $G_{bs\mu}^{R}$, within the $1\sigma$ range
that explains $R_{K^{(*)}}$ according to the latest global fits \cite{Geng:2021nhg}.
This leads to a more involved and highly constrained analysis than often considered, which is summarised as follows.

Explaining both muon $g-2$ and $R_{K^{(*)}}$ anomalies for $M_{Z'}>2\,\mathrm{TeV}$
becomes challenging if we consider the theoretically clean fit where
the positive $1\sigma$ region of $G_{bs\mu}^{R}$ is small. This is because larger values of 
$g_{\mu\mu}^{R}$ are required to explain $\left(g-2\right)_{\mu}$,
but then this implies smaller values of $g_{bb}$ to keep $G_{bs\mu}^{R}$
within the $1\sigma$ range of the theoretically clean fit. For heavier masses of $Z'$,
larger values of $g_{bb}$ are required to fit $G_{bs\mu}^{L}$. However, the heavier the $Z'$
boson is, the larger the values of $M_{4}^{C}$ must be to explain $\left(g-2\right)_{\mu}$, with
$M_{4}^{C}$ bounded from above by perturbation theory, $M_{4}^{C}\apprle\sqrt{4\pi}v/\sqrt{2}\approx618\,\mathrm{GeV}$.
On the other hand, if we consider the global fit that includes angular
observables of $B\rightarrow K^{*}\bar{\mu}\mu$ data, here $G_{bs\mu}^{R}$
is compatible with larger positive values and hence also larger $g_{\mu\mu}^{R}$
are allowed, but the perturbation theory constraint over $M_{4}^{C}$
remains. Despite these challenges we have been able to find viable regions of parameter space which can explain both the 
muon $g-2$ and $R_{K^{(*)}}$, for both global fits.

Finally, we have studied the impact of collider searches for this simplified 
model: constraints coming from experimental measurements of $Z\rightarrow4\mu$
can be avoided by keeping ${M_{Z'}>70\,\mathrm{GeV}}$ \cite{Falkowski:2018dsl,Altmannshofer:2014cfa,Altmannshofer:2014pba,Altmannshofer:2016jzy}.
However, dimuon resonance searches by ATLAS \cite{ATLAS:2017fih}
are already very constraining for $M_{Z'}>500\,\mathrm{GeV}$, in
such a way that the good results of this model for heavy $Z'$ masses
could be probed by the the upcoming LHC run 3 starting in 2022.

\section*{Acknowledgements}
This project has received funding from the European Union's
Horizon 2020 Research and Innovation programme under Marie Sk\l{}odowska-Curie
grant agreement HIDDeN European ITN project (H2020-MSCA-ITN-2019//860881-HIDDeN).
SFK acknowledges the STFC Consolidated Grant ST/L000296/1.

\bibliography{RK_g-2_MFN_SFK.bbl}

\end{document}